\documentclass[12pt]{iopart}
\usepackage{iopams}
\usepackage{epsfig}
\usepackage{array}

\usepackage{longtable}
\usepackage{rotating}

\def\thesection {\Roman{section}}

\begin{document}

\title[Relevance of baseline hard p+p spectra for high-energy A+A physics]
{Relevance of baseline hard proton-proton spectra for high-energy nucleus-nucleus physics}

\author{David~d'Enterria
\footnote[3]{e-mail:denterria@nevis.columbia.edu}
}

\address{Nevis Laboratories, Columbia University\\ 
Irvington, NY 10533, and New York, NY 10027, USA}



\begin{abstract}
We discuss three different cases of hard inclusive spectra
in proton-proton collisions: high $p_T$ single hadron production 
at $\sqrt{s}\approx$ 20 GeV and at $\sqrt{s}$ = 62.4 GeV, 
and direct photon production at $\sqrt{s}$ = 200 GeV; 
with regard to their relevance for the search of Quark Gluon 
Plasma signals in A+A collisions at SPS and RHIC energies.
\end{abstract}


\submitto{\JPG}



\section*{Introduction}

The study of the fundamental theory of the strong interaction, 
Quantum Chromo Dynamics (QCD), in extreme conditions of 
densities and temperatures has attracted much experimental
and theoretical interest in recent years. Experimentally,
the only available means to investigate the {\it many-body} 
chromo-dynamics of a dense and hot system of partons 
involves the use of large atomic nuclei collided at relativistic 
energies. In this context, the primary goal of high-energy nucleus-nucleus
collisions is the creation and study in the laboratory of 
a deconfined state of ``plasma of quarks and gluons'' (QGP) 
predicted by QCD calculations on the lattice~\cite{latt} for values 
of the energy density, $\epsilon\gtrsim$ 0.7 GeV/fm$^3$, 
five times larger than those found in the nuclear ground state.\\


The search of QGP signals in the A+A data 
relies strongly on direct comparison to reference 
results from proton-proton collisions in free space as
they provide the ``QCD vacuum'' baseline to which one compares 
the heavy-ion results in order to extract information about the 
properties of the produced hot and dense QCD medium.
Among all available experimental observables, hard probes
provide the most direct information on the fundamental 
quark and gluon degrees of freedom. Indeed, in all 
hadronic collisions (p+p, p+A or A+A), the production of 
particles with high transverse momentum (jets, single 
hadrons with $p_{T} \gtrsim$ 2 GeV/$c$, prompt $\gamma$)
or large mass (heavy quarks), results from direct 
parton-parton scatterings with large momentum transfer $Q^2$ 
(``hard processes''). Since the hard cross-sections 
can be theoretically calculable by perturbative methods via the 
collinear factorization theorem~\cite{factor},
inclusive high $p_T$ hadrons, jets, direct photons, 
Drell-Yan, and heavy flavors, have long been considered 
both experimentally and theoretically sensitive and well 
calibrated probes of the small-distance QCD phenomena.\\

Simple arguments based on QCD factorization for 
hard cross-sections in A+A collisions (with its implicit 
premise of incoherent parton-parton scattering)~\cite{dde_qm04}, 
and direct experimental measurements of Drell-Yan production
in Pb+Pb at CERN-SPS~\cite{na50_drellyan}, and of 
prompt-$\gamma$~\cite{justin_qm04} and total charm yields~\cite{phnx_AuAucharm} 
in Au+Au at BNL-RHIC, support that hard inclusive cross-sections 
in A+A reactions scale simply as $A^2$ times the corresponding
hard p+p cross-sections: 
$E\,d\sigma_{AA\rightarrow hX}^{hard}/d^3p=A^2\cdot E\,d\sigma_{pp\rightarrow hX}^{hard}/d^3p$.
For a given centrality bin (or impact parameter $b$) 
in a nucleus-nucleus reaction, such a simple scaling rule 
translates into the so-called ``$N_{coll}$ (binary) scaling''
relation between hard p+p cross-sections and A+A yields:
$E\,dN_{AA\rightarrow hX}^{hard}/d^3p\,(b)=\langle T_{AA}(b)\rangle\cdot E\,d\sigma_{pp\rightarrow hX}^{hard}/d^3p$,
where $T_{AA}(b)$ is the Glauber (eikonal) nuclear overlap 
function\footnote{The term ``$N_{coll}$ scaling'' comes from the
fact that the average number of nucleon-nucleon ($NN$) collisions 
at $b$ is simply $\langle N_{coll}(b)\rangle = \langle T_{AA}(b)\rangle\cdot \sigma_{pp}^{inel}$.} 
at $b$. The standard method to quantify the effects of 
the medium in a given hard probe produced in a A+A reaction is, thus,
given by the {\it nuclear modification factor}:
\begin{equation} 
R_{AA}(p_{T},y;b)\,=\frac{\mbox{\small{``hot/dense QCD medium''}}}{\mbox{\small{``QCD vacuum''}}}\,=\,\frac{d^2N_{AA}/dy dp_{T}}{\langle T_{AA}(b)\rangle\,\times\, d^2 \sigma_{pp}/dy dp_{T}},
\label{eq:R_AA}
\end{equation}
which measures the deviation of A+A at 
$b$ from an incoherent superposition of $NN$ collisions.\\

In this paper we discuss experimental results on inclusive 
single hadrons at high $p_T$ in p+p and A+A collisions at top 
CERN-SPS energies ($\sqrt{s}\approx$ 20 GeV) and intermediate 
RHIC energies ($\sqrt{s}$ = 62.4 GeV), as well as theoretical 
expectations on direct photon production in p+p and A+A collisions 
at top RHIC energies ($\sqrt{s}$ = 200 GeV). 
The corresponding high $p_T$ hadron and photon A+A nuclear 
modification factors provide, respectively, critical information 
on established QGP signals such as high $p_T$ leading hadron 
suppression (due to parton energy loss in the deconfined system) 
and thermal photon emission from the plasma.


\section*{\underline{Case I}: High $p_T$ p+p $\rightarrow$ $\pi^0 +X$
reference at $\sqrt{s}\approx$ 20 GeV}

One of the canonical signatures of QGP formation in A+A collisions
is the observation of a suppressed production of high $p_T$ leading
hadrons (as compared to p+p collisions in free space) due to 
the non-Abelian energy loss of the parent parton traversing the dense 
medium produced in the A+A reaction (``jet quenching'')~\cite{jet_quench_review}.
In agreement with these expectations, high-$p_T$ hadroproduction in 
central Au+Au collisions at $\sqrt{s_{\mbox{\tiny{\it{NN}}}}}$ = 130, 200 GeV at 
RHIC has been found to be strongly suppressed (by up to a factor of 4--5)
~\cite{phenix_hipt_130,star_hipt_130,phenix_hipt_pi0_200,star_hipt_200,phobos_hipt_200,brahms_hipt_200} 
compared to p+p collisions measured at the same $\sqrt{s_{\mbox{\tiny{\it{NN}}}}}$
~\cite{phenix_pp_pi0_200,star_hipt_200}. 
In contrast to this result, high $p_T$ pion production in 
central A+A at CERN-SPS energies was found not to be suppressed but 
{\it enhanced} compared to production in free space~\cite{wa98,wang_sps,wang_syst}.
Such a ``Cronin effect'' was consistent with previous observations 
in fixed-target p+A collisions at Fermilab
($\sqrt{s_{\mbox{\tiny{\it{NN}}}}}\approx$ 20 -- 40 GeV)~\cite{cronin,antrea,straub} 
and in $\alpha+\alpha$ interactions at the ISR collider
($\sqrt{s_{\mbox{\tiny{\it{NN}}}}}$ = 31 GeV)~\cite{isr_alpha2_pi0}. 
The prevalence of the Cronin broadening, characteristic ``cold QCD matter'' 
systems, in A+A reactions at the SPS suggested that multiple initial-state (soft) 
$k_T$ ``kicks'' suffered by the colliding parton in their way through 
the nucleus dominated over possible final-state energy-loss effects for
center-of-mass energies of order $\sqrt{s_{\mbox{\tiny{\it{NN}}}}}\approx$ 20 GeV.\\

A recent reanalysis of the SPS results~\cite{dde_hipt_SPS} 
has shown, however, that the apparent strong pion enhancement observed 
is not actually supported by the A+A data but is due to the use of 
inexact p+p baseline references. Indeed, since no concurrent 
measurement of high $p_T$ pion production in proton-proton collisions
was carried out at SPS at the same $\sqrt{s}$ as the heavy-ion
experiments (and given that perturbative QCD calculations are not
reliable in this relatively low energy domain) one had to count on 
p+p parametrizations extrapolated from data at higher collision energies. 
On the one hand, the WA98 collaboration~\cite{wa98} employed a empirical 
power-law form $A\,[p_0/(p_T+p_0)]^n$ (originally proposed by
Hagedorn~\cite{hagedorn}) tuned to reproduce the ISR p+p pion spectra 
($\sqrt{s}\approx$ 20 -- 31 GeV), plus an $x_T$ scaling 
prescription~\cite{beier} to account for the collision energy dependence 
of the cross section. On the other hand, Wang\&Wang~\cite{wang_syst} adopted a 
more complex power-law ansatz for the $p_T$ spectrum which described the 
charged pion data at $\sqrt{s}$ = 19.4 GeV~\cite{antrea}, combined with a pQCD 
parton model calculation to scale the cross-section down to $\sqrt{s}$ = 17.3 GeV.
Though both parametrizations were tuned to reproduce a subset of the 
existing p+p $\rightarrow \pi$+$X$ data at $\sqrt{s}\approx$ 20 GeV, no
true global analysis was carried out to fully compare the parametrizations 
to all the existing data in this energy regime. 

\begin{figure}[htbp]
\begin{center}
\begin{minipage}[c]{.44\linewidth}
\includegraphics[height=8.0cm,width=8.0cm]{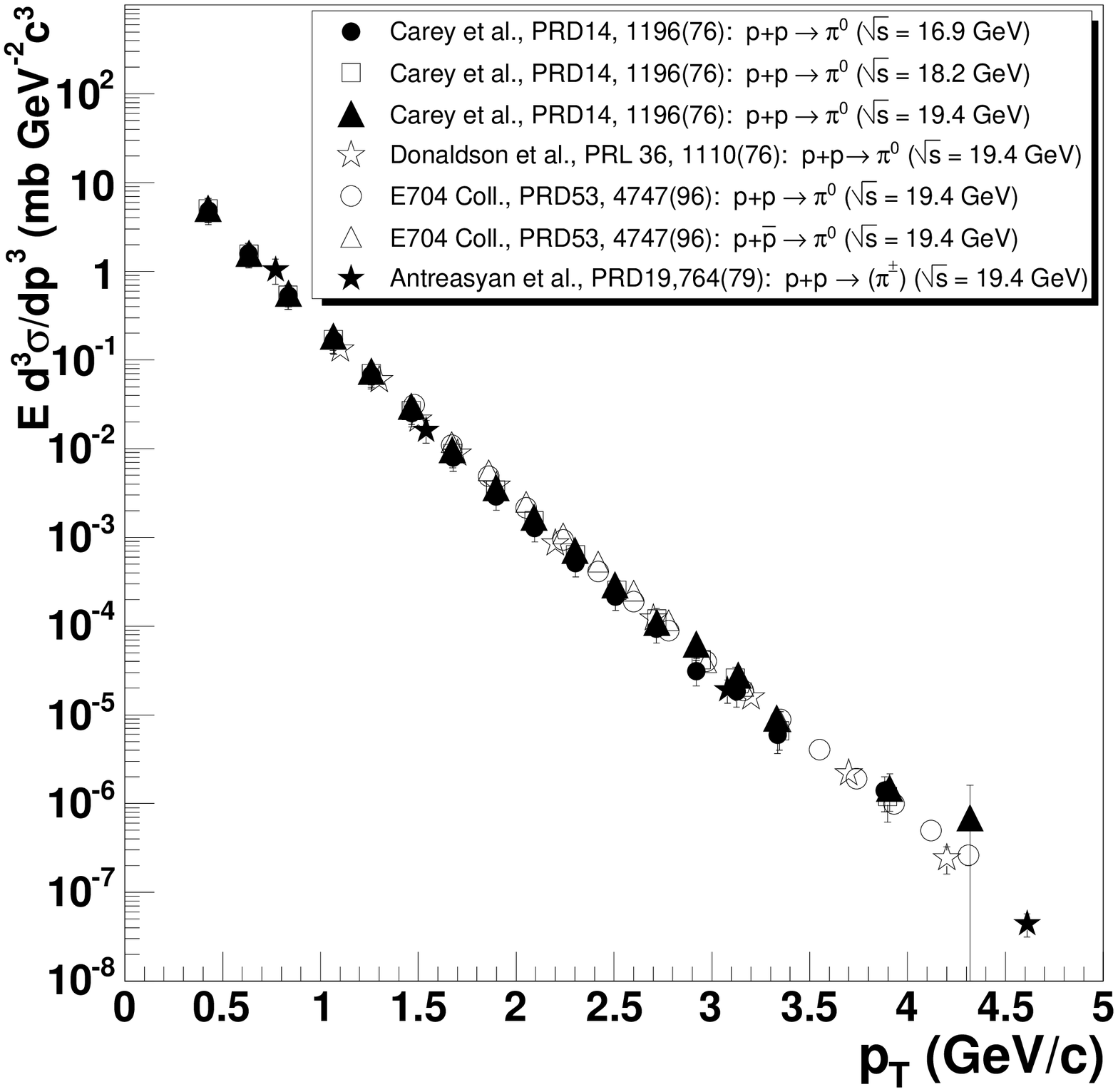}\hfill
\end{minipage} \hfill
\begin{minipage}[c]{.47\linewidth}
\includegraphics[height=5.50cm,width=7.50cm]{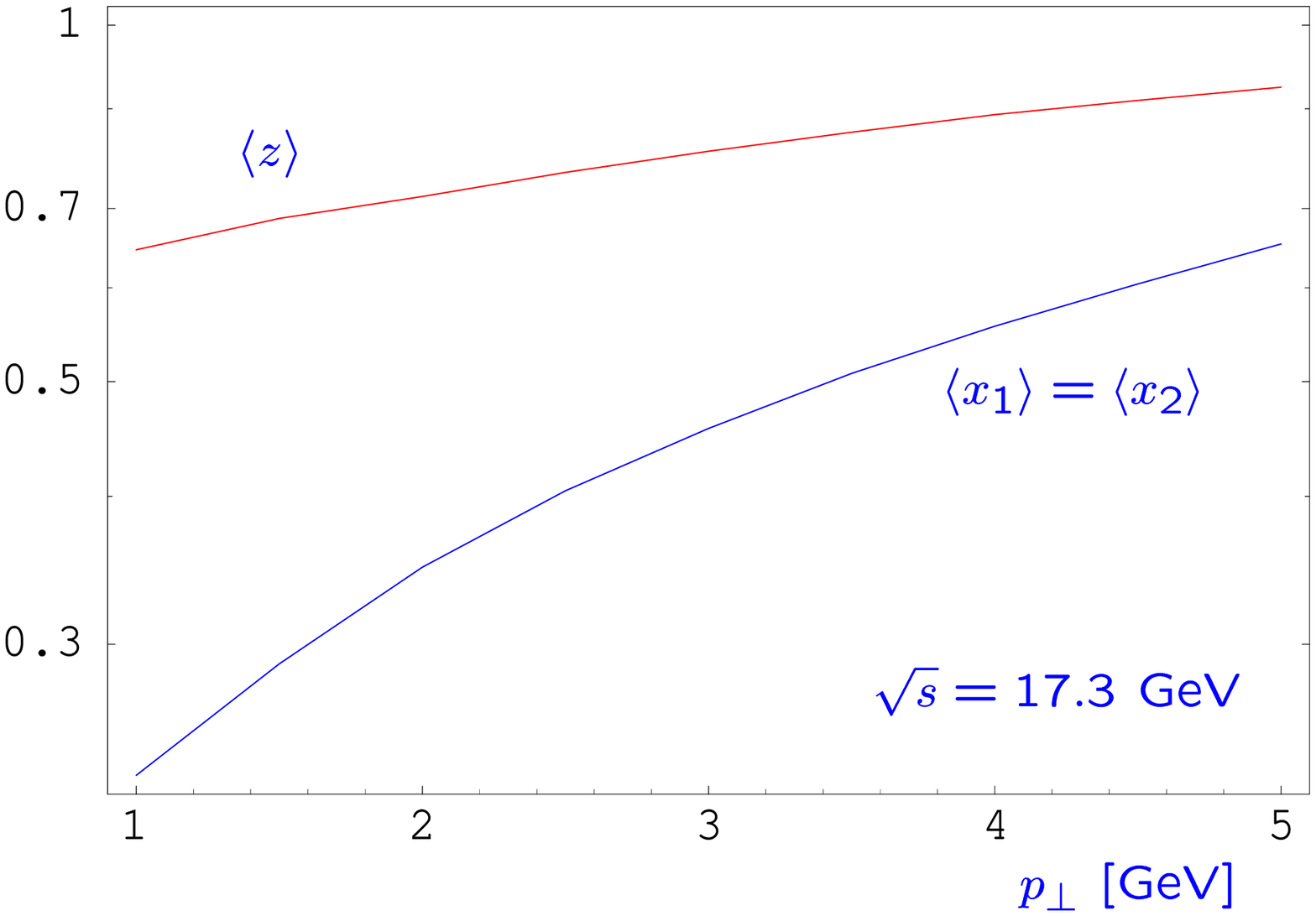}\hfill
\end{minipage} \hfill
\end{center}
\caption{Left: Compilation of single inclusive pion spectra measured at high $p_T$
($p_{T} \gtrsim$ 2 GeV/$c$) in p+p collisions~\protect\cite{antrea,carey,donaldson,adams} 
in the same collision energy range ($\sqrt{s}$ = 16.9 -- 19.4 GeV)
as the CERN-SPS A+A experiments.
Right: Scaling variables $\langle x_{1,2} \rangle$ (average parton fractional
momentum) and $\langle z \rangle$ (average momentum fraction of the parent parton
carried by the leading pion) for p+p $\rightarrow\pi^0$ production ($\sqrt{s}$ = 17.3 GeV)
at mid-rapidity versus the $\pi^0$ momentum, computed in perturbative QCD~\protect\cite{kretzer0}.}
\label{fig:pp_pi0_spec}
\end{figure}

\begin{figure}[htbp] 
\begin{center}
\epsfig{file=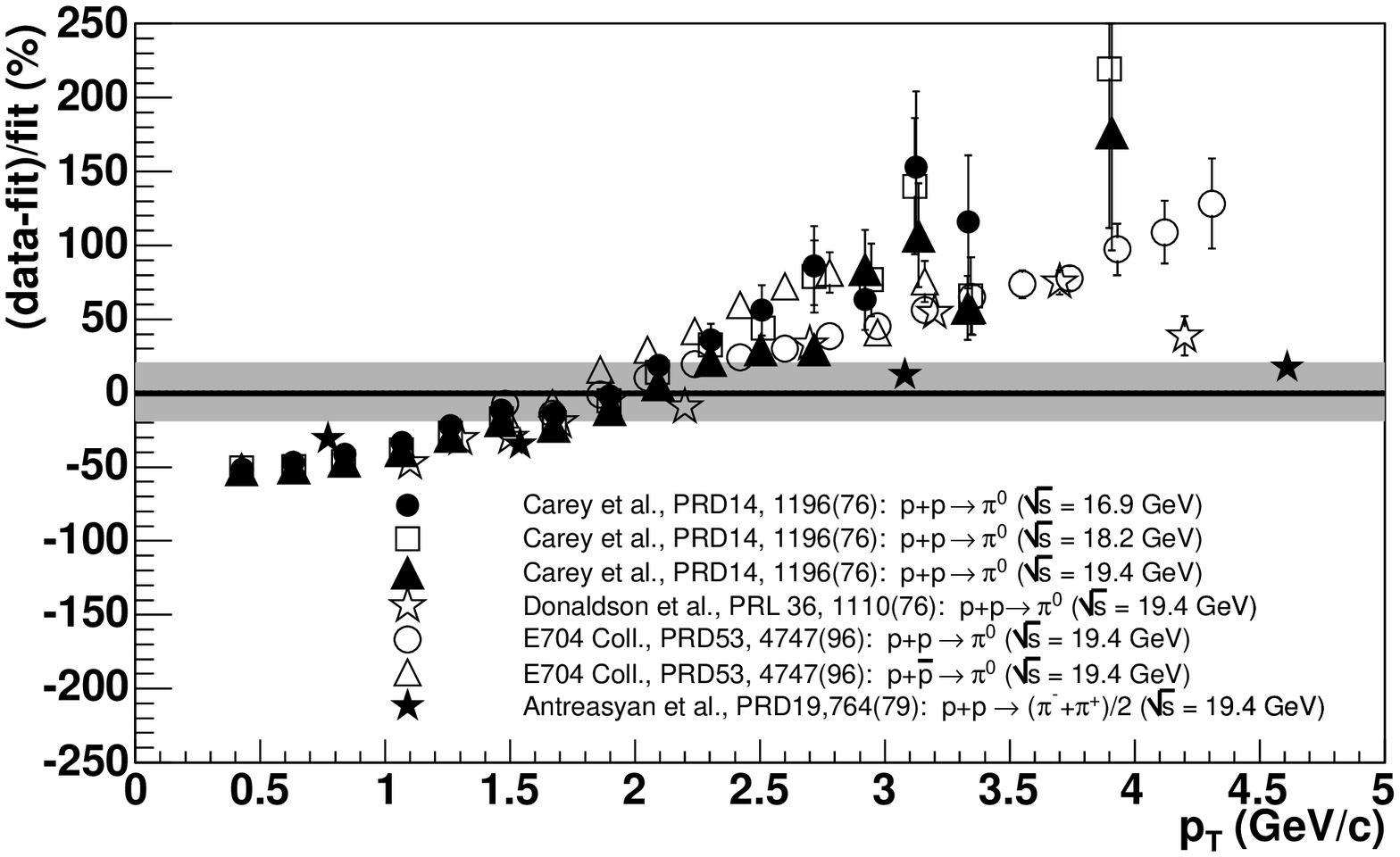,height=6.25cm,width=11.5cm}
\epsfig{file=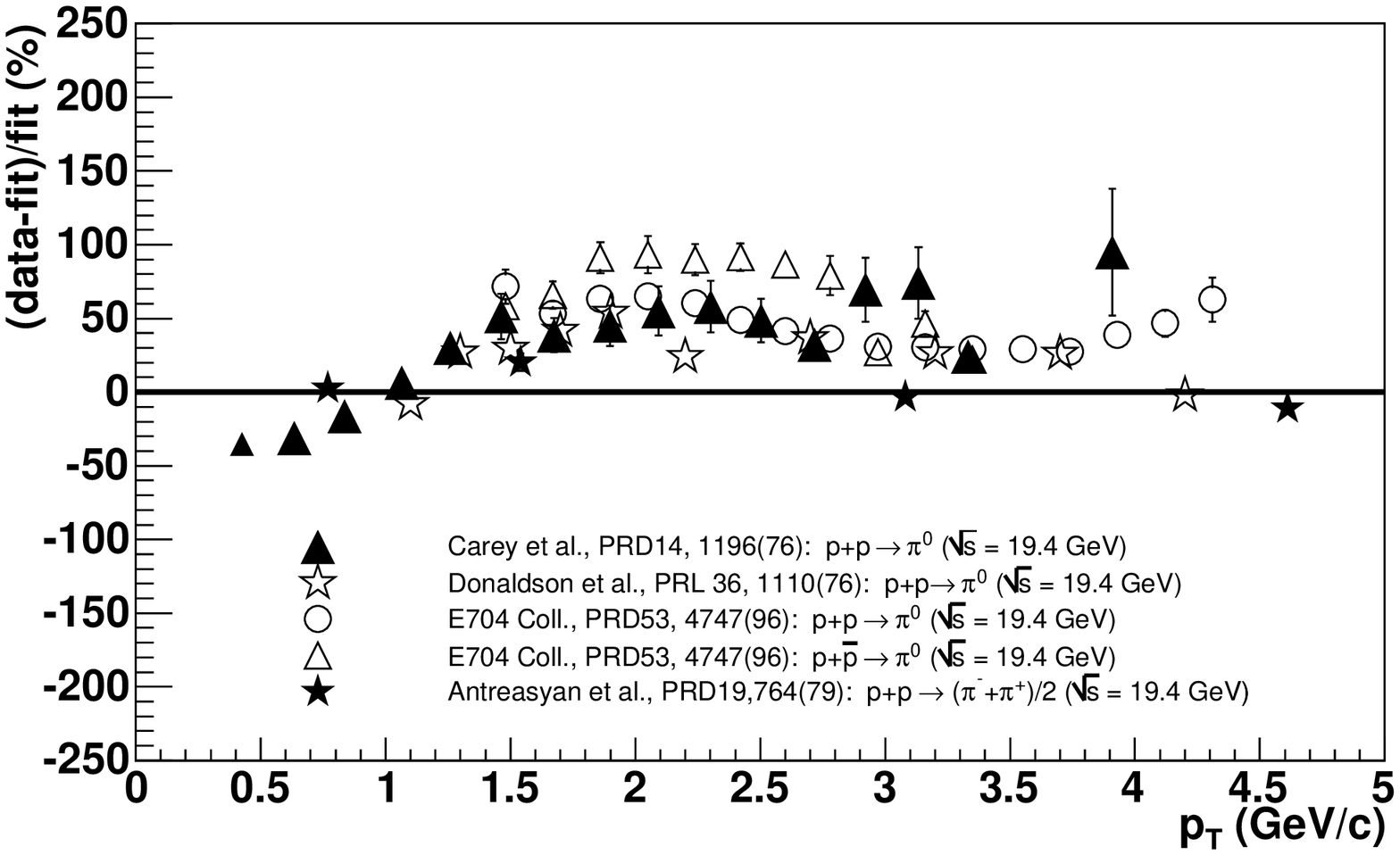,height=6.25cm,width=11.5cm}
\end{center}
\caption{Relative differences between the single inclusive pion 
spectra measured in p+p collisions at $\sqrt{s}$ = 16.9 -- 19.4 
GeV~\protect\cite{antrea,carey,donaldson,adams} 
and the p+p $\rightarrow \pi^0$+$X$ parametrizations proposed 
by the WA98 collaboration~\protect\cite{wa98} (upper figure) 
and Wang\&Wang~\protect\cite{wang_syst} (lower figure) at the 
corresponding $\sqrt{s}$. The shaded band represents the 20\% 
overall uncertainty assigned originally to the WA98 parametrization.
Wang\&Wang only provides the fit parameters for 2 fixed values 
of $\sqrt{s}$ = 17.3, 19.4~GeV.}
\label{fig:wa98_xnwang_fits}
\end{figure}

Figure~\ref{fig:pp_pi0_spec} (left) shows all single inclusive pion spectra measured 
at high $p_T$ in p+p,$\bar{\mbox{p}}$ collisions in the same $\sqrt{s}$ range of 
the CERN-SPS heavy-ion experiments, whereas the plot in Figure~\ref{fig:pp_pi0_spec} (right)
shows the relevant parton kinematical domain for hard hadro-production in this energy 
regime. Fig.~\ref{fig:wa98_xnwang_fits} confronts the two proposed parametrizations 
to the experimental data. Both parametrizations fail to describe adequately the shape 
of the $p_T$ spectra, and both undershoot the cross-sections\footnote{Note that, 
as discussed in~\cite{dde_hipt_SPS}, the $\pi^0$ data of ~\cite{carey} plotted 
in Figs.~\ref{fig:pp_pi0_spec}--\ref{fig:wa98_xnwang_fits} 
have been even scaled {\it down} by a factor of 0.8 above $p_T$ = 1.5 GeV/$c$ to 
take into account the $\eta$ contamination not subtracted in the original 
tabulated results (see a discussion of this effect in Section~\ref{sec:eta_correction}).} 
by as much as by factors of  2--3 of the same order 
as the reported Cronin enhancements.

\begin{figure}[htbp]
\begin{center}
\epsfig{file=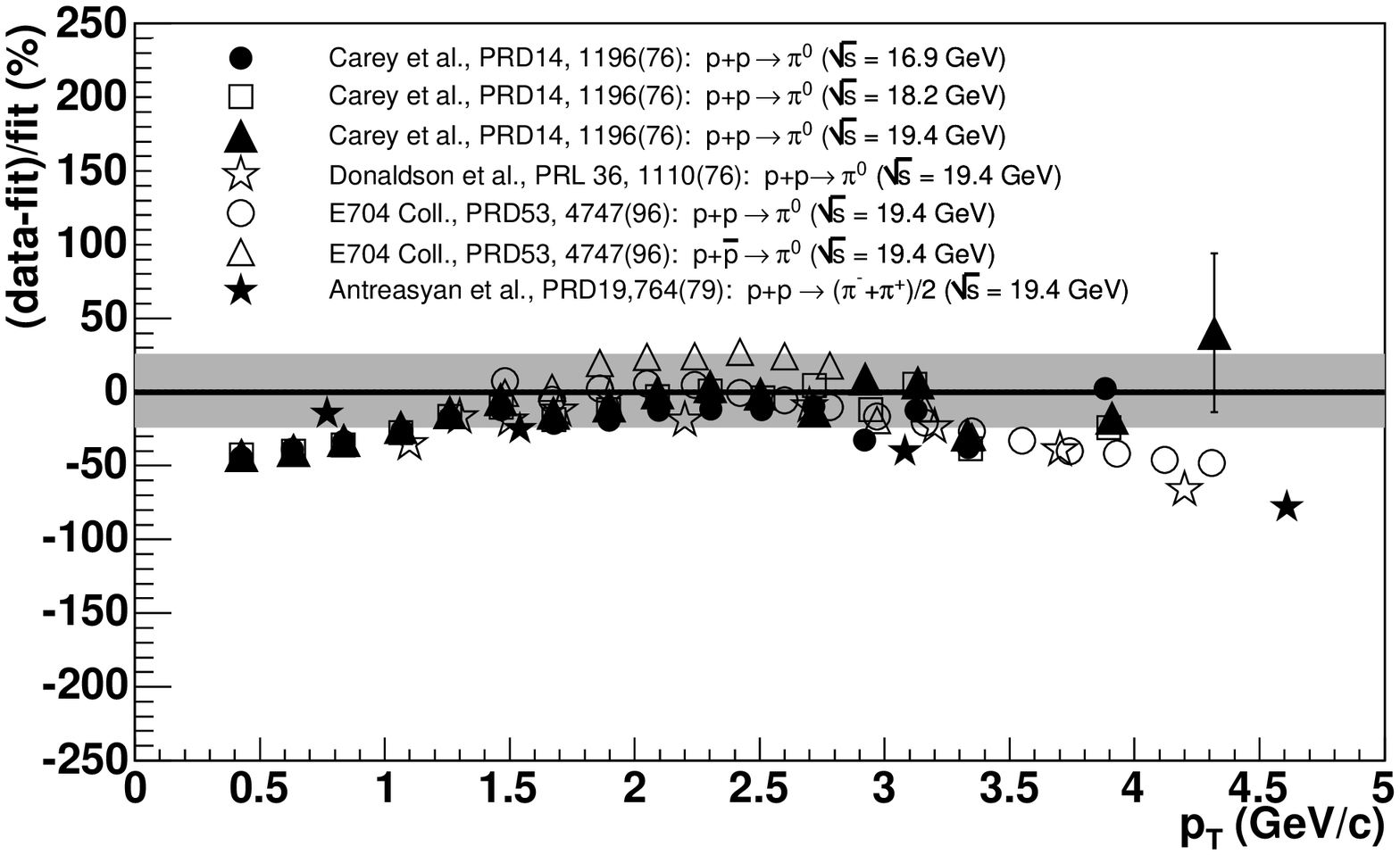,height=6.8cm,width=11.5cm}
\end{center}
\caption{Relative differences between the single inclusive pion 
spectra measured in p+p collisions at $\sqrt{s}$ = 16.9 -- 19.4 
GeV~\protect\cite{antrea,carey,donaldson,adams} and the 
p+p $\rightarrow \pi^0$+$X$ parametrization derived in 
ref.~\protect\cite{blatt}. The shaded band represents the 25\% 
overall uncertainty that we assign to the parametrization.}
\label{fig:fit_blatt}
\end{figure}

As an alternative to the WA98 and Wang\&Wang p+p references, 
we proposed~\cite{dde_hipt_SPS} to employ a purely empirical 
11-parameter functional form derived by Blattnig {\it et. al}~\cite{blatt} 
from a global analysis of most of the available proton-proton
$\pi^0$ spectra measured in the range $\sqrt{s}\approx$ 7 -- 63 GeV. 
Fig.~\ref{fig:fit_blatt} shows that the level of agreement of the Blattnig 
parametrization to the p+p pion cross-sections of Fig.~\ref{fig:wa98_xnwang_fits}.
is more satisfactory, both in shape and magnitude, 
than the two previous parametrizations at least within the $p_T$ range 
($p_T\approx$ 1.5 -- 3.5 GeV/$c$) covered by the heavy-ion experiments. 
Importantly, the p+p parametrization describes reasonable well the 
$\sqrt{s}$ dependence of the yields which is a critical requirement for 
using it as a fair baseline for the  Pb+Pb and Pb+Au data at 
$\sqrt{s_{\mbox{\tiny{\it{NN}}}}}$ = 17.3 GeV.

\begin{figure}[htbp]
\begin{center}
\epsfig{file=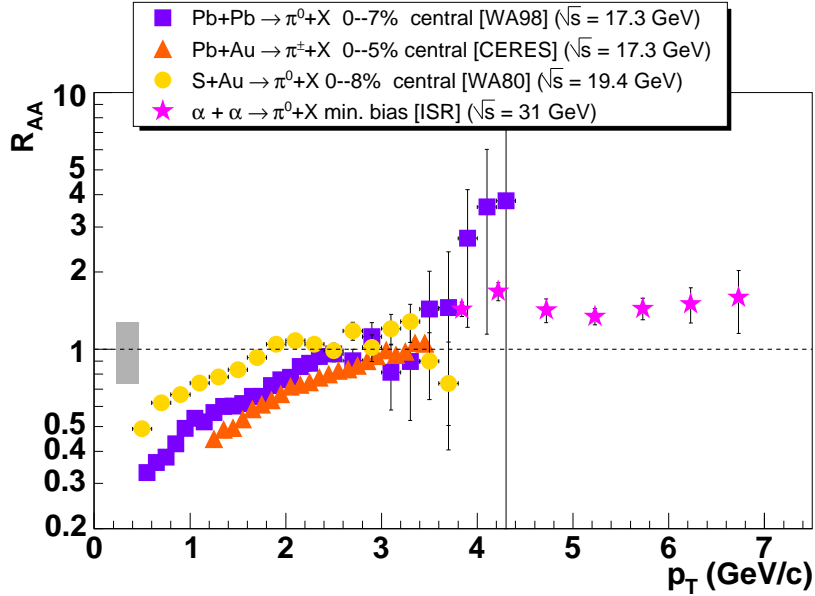,height=8.cm}
\end{center}
\caption{Nuclear modification factors for pion production at CERN-SPS 
in central Pb+Pb~\protect\cite{wa98}, Pb+Au~\protect\cite{ceres},
and S+Au~\protect\cite{wa80} reactions at 
$\sqrt{s_{\mbox{\tiny{\it{NN}}}}}\approx$ 20 GeV, and at ISR in minimum bias
$\alpha+\alpha$ reactions at $\sqrt{s_{\mbox{\tiny{\it{NN}}}}}$ = 31 
GeV~\protect\cite{isr_alpha2_pi0}. The $R_{AA}$ from SPS are obtained using the 
p+p parametrization proposed in ref.~\protect\cite{dde_hipt_SPS}. 
The $R_{AA}$ for ISR $\alpha+\alpha$ has been obtained using the p+p 
spectrum measured in the same experiment~\protect\cite{isr_alpha2_pi0}.
The shaded band at $R_{AA}$ = 1 represents the overall fractional uncertainty 
(including in quadrature the 25\% uncertainty of the p+p reference 
and the 10\% error of the Glauber calculation of $N_{coll}$). 
CERES data~\protect\cite{ceres} have an additional overall uncertainty of 
$\pm$15\% not shown in the plot.}
\label{fig:RAA_SPS_ISR}
\end{figure}

\begin{figure}[htbp]
\centerline{\psfig{figure=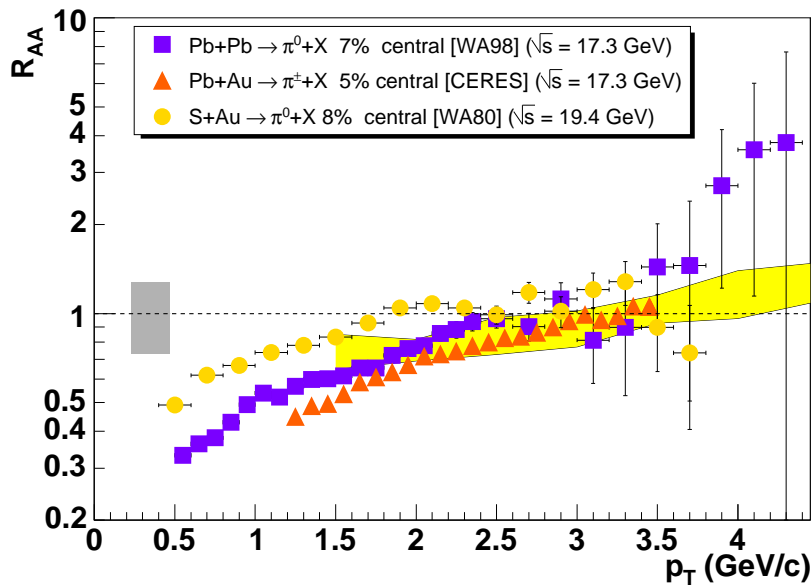,height=8.cm}}
\caption{Nuclear modification factors for pion production at CERN-SPS in 
central Pb+Pb~\protect\cite{wa98}, Pb+Au~\protect\cite{ceres}, 
and S+Au~\protect\cite{wa80} reactions (Fig.~\protect\ref{fig:RAA_SPS_ISR})  
compared to a theoretical prediction from Vitev and 
Gyulassy~\protect\cite{vitev_gyulassy} including standard nuclear effects 
(Cronin broadening and shadowing) and final-state parton energy 
loss in a system with initial gluon densities $dN^g/dy=$ 400 -- 600.}
\label{fig:RAA_centSPS_vs_vitev}
\end{figure}

Using the reference of Blattnig {\it et. al}, we revised in~\cite{dde_hipt_SPS} 
the nuclear modifications factors for the whole set of high $p_T$ data from 
the SPS heavy-ion programme: $\pi^0$ and $\pi^{\pm}$ at 
$\sqrt{s_{\mbox{\tiny{\it{NN}}}}}$ = 17.3 GeV from Pb+Pb (WA98)~\cite{wa98} 
and Pb+Au (CERES/NA45)~\cite{ceres} respectively, and $\pi^0$ from S+Au at 
$\sqrt{s_{\mbox{\tiny{\it{NN}}}}}$ = 19.4 GeV (WA80)~\cite{wa80}. The
revisited $R_{AA}$, plotted in Figure~\ref{fig:RAA_SPS_ISR}, indicate that
high-$p_T$ hadroproduction at $\sqrt{s_{\mbox{\tiny{\it{NN}}}}}\approx$ 20 GeV 
is not enhanced in central nucleus-nucleus reactions but, within errors, 
is consistent instead with scaling with the number of $NN$ collisions.
Interestingly, at variance with central collisions, high $p_T$ pion production 
in {\it peripheral} Pb+Pb reactions at SPS is indeed enhanced by as much as by a factor 
of $\sim$2 compared to the ``$N_{coll}$ scaling'' expectation~\cite{wa98,dde_hipt_SPS}.
This fact indicates that there must exist a mechanism in central A+A that 
``neutralizes'' the Cronin enhancement down to values consistent with $R_{AA}\approx$ 1.
Theoretical predictions~\cite{vitev_gyulassy} of high $p_T$ $\pi^0$ production 
in central Pb+Pb collisions including Cronin broadening and (anti)shadowing
supplemented with final-state partonic energy loss in an expanding dense system 
with effective gluon densities $dN^g/dy=$ 400 -- 600 reproduce well the observed 
suppression factor (yellow band in Fig.~\ref{fig:RAA_centSPS_vs_vitev}) 
supporting the idea that a moderate amount of jet quenching is already 
present in the most central heavy-ion reactions at SPS. The confirmation
of this conclusion requires, however, a direct (and accurate) measurement
of the high $p_T$ p+p pion cross-section at $\sqrt{s}$ = 17.3 GeV.
Unfortunately, although RHIC can run Au+Au at a similar center-of-mass energy
of $\sqrt{s_{\mbox{\tiny{\it{NN}}}}}$ = 19.6 GeV (corresponding to the 
9 GeV injection energy from AGS), the minimum collision energy 
in the proton-proton mode is $\sqrt{s}$ = 48.6 GeV (the RHIC design injection 
energy for each proton beam from AGS is 24 GeV, above the transition 
energy of 22 GeV)~\cite{adrees}. This fact hinders the possibility of
directly comparing high $p_T$ hadroproduction in A+A and p+p
collisions at RHIC at center-of-mass energies comparable to SPS.

\section*{\underline{Case II}: High $p_T$ p+p $\rightarrow$ $\pi^0$, $h^\pm +X$ 
references at $\sqrt{s}$ = 62.4 GeV.}

The study of the excitation function of high $p_T$ hadron suppression between 
top SPS and top RHIC energies was the main motivation behind the dedicated 
Au+Au run at RHIC intermediate energies ($\sqrt{s_{\mbox{\tiny{\it{NN}}}}}$ = 62.4 GeV)
carried out in April 2004.
PHENIX measured charged hadrons up to $p_T$ = 5 GeV/$c$ and identified 
neutral pions in the range $p_T$ = 1 -- 7 GeV/$c$~\cite{phenix_hipt_62}.
PHOBOS~\cite{phobos_hipt_62} and STAR~\cite{star_hipt_62} measured inclusive 
charged hadrons up to $p_T\approx$ 4.5 GeV/$c$ and 12 GeV/$c$ respectively. 
However, as in the SPS case, no concurrent p+p reference measurement was performed
at $\sqrt{s}$ = 62.4 GeV, and the corresponding Au+Au nuclear modifications factors 
were constructed using, basically, p+p $\rightarrow$ $h^\pm,\pi^0$ differential 
cross-sections measured in the 70s and 80s at the top ISR energies 
($\sqrt{s}$ = 62 $-$ 63 GeV). 
Table~\ref{tab:compilation} collects all the existing measurements of 
high $p_T$ (neutral and charged) pion and inclusive charged 
hadrons in proton-proton collisions at the maximum CERN-ISR energy. 




\begin{sidewaystable}[htbp]
\begin{scriptsize}
\begin{center} 
\setlength{\extrarowheight}{0.2cm}
\begin{tabular}{|c|c|c|c|c|c|c|c|c|c|c|c|}\hline\hline
Reaction  &  $\sqrt{s}$  & Collab./Exp. & Ref. & $y$ & $p_T$ range  & Data & Energy scale & Extra syst. \& & $\sqrt{s}$ & direct-$\gamma$ & $\eta$ \\
 & (GeV) &  &  &  & (GeV/$c$) & points &  error in yield (\%) & abs.norm. error & correction & correction & correction \\\hline
p+p $\rightarrow \pi^{0}+ X$ & 62.4 & CCR & (busser73)~\cite{busser73,busser74} & 0.0 & 2.9 -- 4.6 & 7 & 55\% (6\%) & -- & not needed & needed & needed \\
p+p $\rightarrow \pi^{0}+ X$ & 62.9 & ACHM & (eggert75)~\cite{eggert75} & 0.0 & 0.7 -- 6.4 & 29 & 35\% (5\%) & 5\% & needed & needed & needed \\
p+p $\rightarrow \pi^{0}+ X$ & 62.4 & CCRS & (busser76)~\cite{busser76} & 0.0 & 2.4 -- 6.2 & 40 &  26\% (3\%) & -- & not needed & needed & not needed \\
p+p $\rightarrow \pi^{0}+ X$ & 63   & CSZ & (clark78)~\cite{clark78} & 0.0 & 5.2 -- 16.5 & 17 & 25\% (3\%$\oplus$2\%) & 17\% & needed & needed & needed \\
p+p $\rightarrow \pi^{0}+ X$ & 62.4 & CCOR/R-108 & (angelis78)~\cite{angelis78} & 0.0 & 3.7 -- 13.7 & 21 & 25\% (5\%) & 5\% & not needed & needed & needed \\
p+p $\rightarrow \pi^{0}+ X$ & 62.4 & R-806 & (kourkou79)~\cite{kourkou79} & 0.0 & 3.0 -- 15.0 & 28 & 22\% & -- & not needed & needed & not needed \\
p+p $\rightarrow \pi^{0}+ X$ & 62.8 & R-806 & (kourkou80)~\cite{kourkou80} & 0.0 & 3.0 -- 15.0 & 41 & 35\% (*) & -- & needed & needed & not needed \\
p+p $\rightarrow m^{0}+ X^{\ddagger}$ & 62.4  & CMOR & (angelis89)~\cite{angelis89} & 0.0 & 4.7 -- 9.0 & 7 & 25\% (5\%) & 17\% & not needed & not needed & needed \\
p+p $\rightarrow \pi^{0}+ X$ & 63   & AFS & (akesson89)~\cite{akesson89} & 0.0 & 4.7 -- 13.7 & 11 & -- (**) & -- & needed & not needed & not needed \\ \hline\\\hline

p+p $\rightarrow \pi^{\pm}+ X$ & 63 & Brit.-Scand. & (alper75)~\cite{alper75} & 0.0 & 0.1 -- 2.4 & 17 &  -- & -- & not needed$^\dagger$ & -- & --\\ 
p+p $\rightarrow \pi^{\pm}+ X$ & 62 & CCRS & (busser76)~\cite{busser76}   & 0.0 & 3.3 -- 8.0 & 22 & -- & -- & not needed & -- & -- \\
p+p $\rightarrow \pi^{\pm}+ X$ & 62 & Saclay & (banner77)~\cite{banner77} & 0.0 & 0.2 -- 1.5 & 21 & -- & -- & not needed & -- & --\\
p+p $\rightarrow \pi^{\pm}+ X$ & 63 & SFM & (drijard82)~\cite{drijard82} & 0.8 & 3.8 -- 12.5 & 21 & -- & -- & needed & -- & --\\
\hline\\\hline

p+p $\rightarrow h^{\pm}+ X$ & 63 & AFS  & (akesson82)~\cite{akesson82} & 0.0  & 2.25 -- 5.8 & 10 & -- & -- & not needed$^\dagger$ & -- & --\\
p+p $\rightarrow h^{\pm}+ X$ & 63 & CDHW & (breakst95)~\cite{breakstone95} & 0.25 & 0.25 -- 3.0 & 11 & -- & -- & not needed$^\dagger$ & -- & --\\
p+p $\rightarrow h^{\pm}+ X$ & 63 & CDHW & (breakst95)~\cite{breakstone95} & 0.75 & 0.25 -- 3.0 & 11 & -- & -- & not needed$^\dagger$ & -- & --\\ 
\hline 
\hline 
\end{tabular}\vspace{3mm} 
\caption{Chronological compilation of $\pi^{0,\pm}$ [$^{\ddagger}\; m^0= \pi^0 + \eta$], 
and inclusive charged ($h^\pm$) production measured at the top CERN-ISR energies. 
For each (1) reaction, we quote the (2) center-of-mass energy, (3) collaboration/experiment 
name, (4) bibliographical reference, (5) rapidity domain, (6) measured $p_T$ range 
(center of min. and max. bins quoted), (7) total number of data 
points, (8) $\pi^0$ energy-scale errors (the value in parenthesis is the true 
energy-scale uncertainty and the uncertainty quoted is the total effect of the
error propagated to the yields), (9) additional systematic and/or 
absolute normalization (luminosity) errors [(*) Kourkoumelis {\it et al.} 
energy scale error includes all syst. uncertainties and is an average over $p_T$ 
(error values quoted in the paper are in the range 27\%--42\%). (**) Akesson {\it et al.} 
original point-to-point errors include all uncertainties.]. Column (10) 
lists those measurements that have been corrected to account for the
(slightly) different ISR ($\sqrt{s}$ = 62. -- 63. GeV) and RHIC 
($\sqrt{s}$ = 62.4 GeV) center-of-mass energies (those data sets marked
with a dagger ($^{\dagger}$) have not been revised since they cover 
low $p_T$ values where the effect of the correction on the yields is minimal).
Columns (11) and (12) indicate whether the published $\pi^0$ spectrum has been 
corrected for direct-$\gamma$ and $\eta$ ``contaminations'' resp. as described
in the text.}
\label{tab:compilation} 
\end{center} 
\end{scriptsize}
\end{sidewaystable}


\setcounter{section}{2}
\subsection{p+p $\rightarrow$ $h^\pm +X$ reference at $\sqrt{s}$ = 62.4 GeV}
\label{sec:pp_h_ref_62GeV}

\begin{figure}[htbp]
\begin{center}
\includegraphics[height=8.0cm,width=11.0cm]{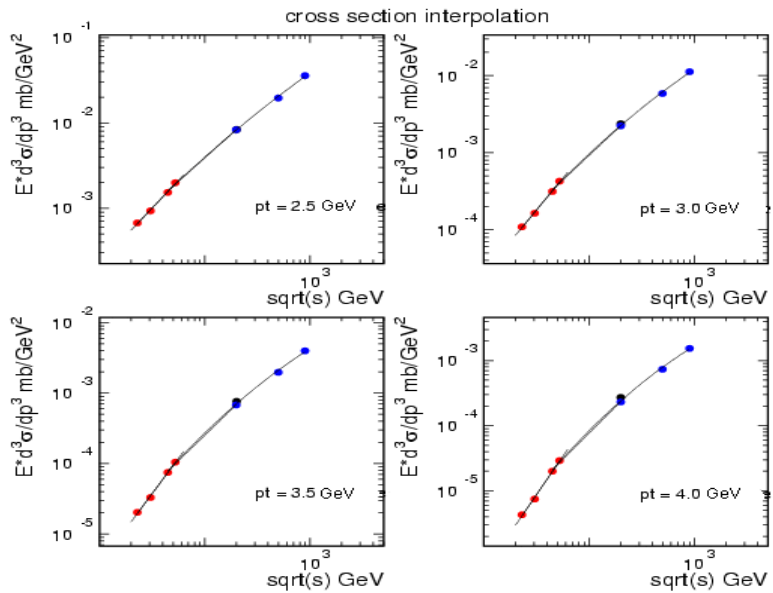}\hfill
\end{center}
\caption{Charged hadron cross-sections as function of center of mass energy for 
fixed $p_T$ values from 2.5 to 4 GeV/$c$~\protect\cite{axel}. The lower energy points (red) are from fits to 
ISR data (combined $\pi^\pm$, $K$ and proton spectra from Alper {\it et al.}~\protect\cite{alper75}), 
the higher energy points (blue) are from fits to UA1 data~\cite{albajar90}, and the black points are from 
a fit to PHENIX data.}
\label{fig:sigma_interpolation_62GeV}
\end{figure}

\begin{figure}[htbp]
\begin{center}
\includegraphics[height=9.5cm,width=9.6cm]{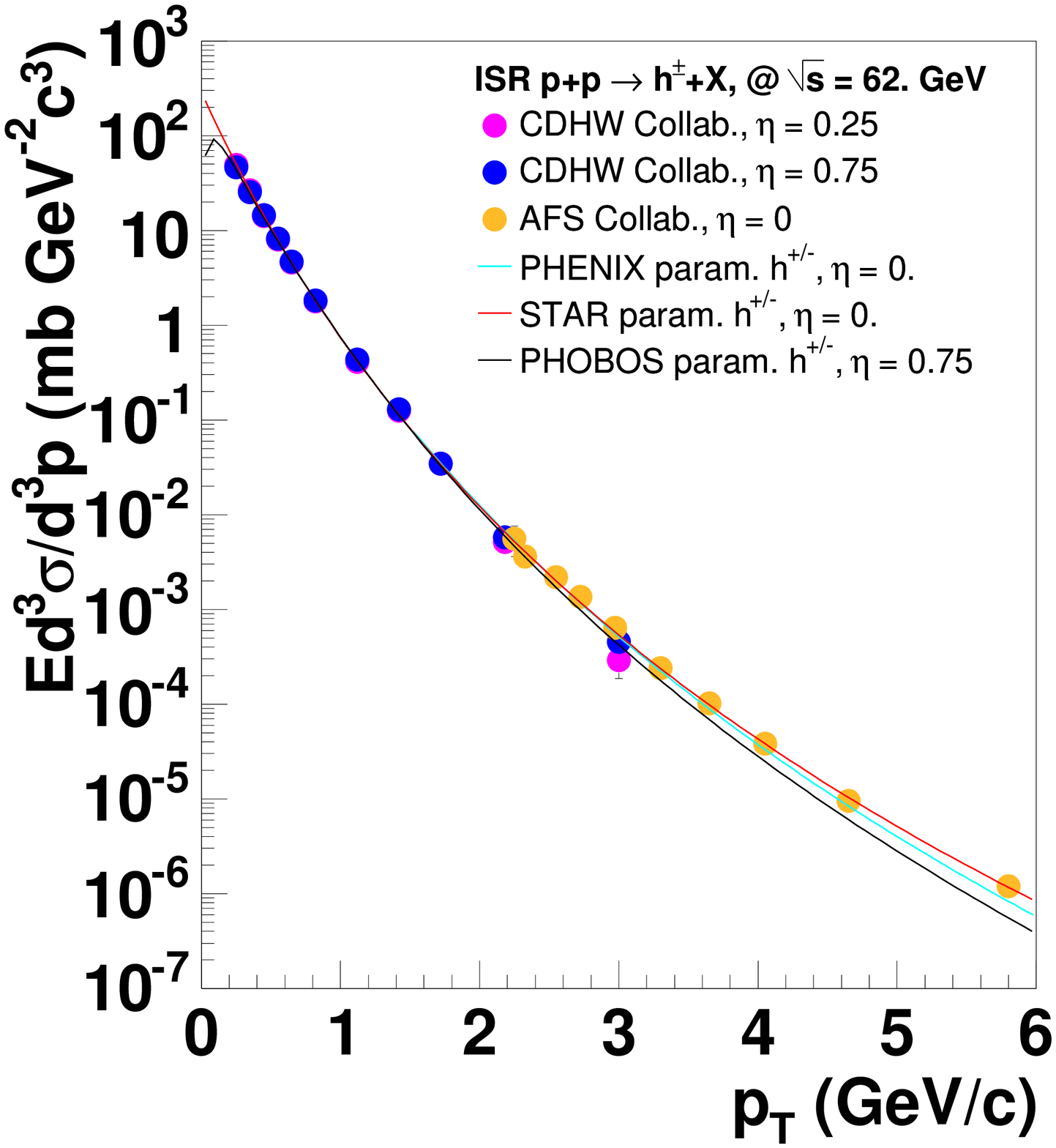}
\includegraphics[height=6.7cm,width=10.cm]{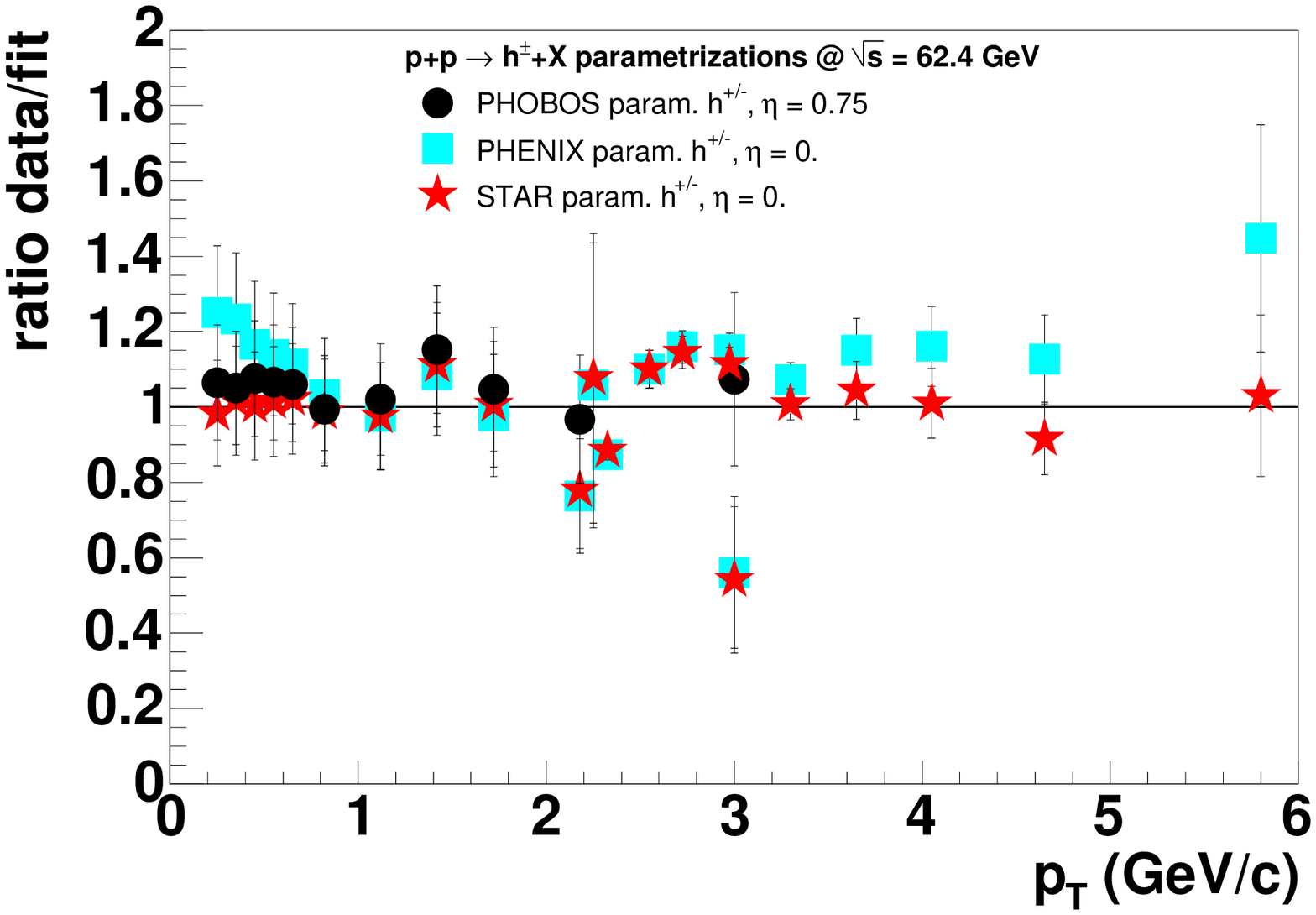}
\end{center}
\caption{Upper: Inclusive charged hadron spectra measured in p+p collisions at 
$\sqrt{s}$ = 63 GeV at CERN-ISR~\cite{akesson82,breakstone95}
at midrapidity ($\eta=$ 0, 0.25) and moderately forward rapidities ($\eta$ = 0.75) 
compared to PHENIX~\cite{axel}, PHOBOS~\cite{phobos_hipt_62} and STAR~\cite{star_hipt_62} 
$h^\pm$ parametrizations. 
Lower: Ratios of the same experimental spectra over the three parametrizations 
(each one evaluated in their valid range of rapidities).}
\label{fig:pp_h_params_62GeV}
\end{figure}

Rather than using the existing ISR $h^\pm$ data at $\sqrt{s}$ = 62 $-$ 63 GeV,
the PHENIX p+p inclusive charged hadron reference at $\sqrt{s}$ = 62.4 GeV has 
been independently obtained~\cite{axel} by interpolating from lower and higher collision 
energy measurements\footnote{The same procedure was followed 
to obtain the p+p reference for RHIC Run-1 Au+Au measurement at 
$\sqrt{s}$ = 130 GeV~\cite{phenix_hipt_130}.}:
p+p at $\sqrt{s}$ = 21, 31, 44, and 53 GeV at CERN-ISR~\cite{alper75}, 
PHENIX p+p at 200 GeV, and CERN-UA1 p+$\bar{\mbox{p}}$
data at 200, 500, and 900 GeV~\cite{albajar90}.
Fig.~\ref{fig:sigma_interpolation_62GeV} shows an example of the cross-section 
interpolation procedure in four individual $p_T$ bins from
the lower and higher $\sqrt{s}$ measurements. The resulting interpolated 
spectrum at 62.4 GeV is fitted to a modified power-law form:
$Ed^3\sigma_{pp\rightarrow hX}/d^3p = A\,(e^{a\cdot x}+p_T/p_0)^{-n}$, with {\it preliminary} 
parameters $A$ = 196.4 [mb GeV$^{-2}c^{3}$], $a$ =  0.0226 [GeV$^{-1}c$], 
$p_0$ = 2.301 [GeV/$c$], and $n$  = 14.86; with an assigned overall uncertainty
of $\pm$25\%~\cite{axel}. As an independent cross-check the fit is compared (Fig.~\ref{fig:pp_h_params_62GeV})
to the three ISR $h^\pm$ data sets measured at 63 GeV (three bottom 
rows of Table~\ref{tab:compilation}). The agreement data--fit is good within 
the assigned $\pm$25\% uncertainty of the parametrization.\\

PHOBOS~\cite{phobos_hipt_62} fitted the experimental p+p data from the CDHW experiment~\cite{breakstone95}
at $\eta$ = 0.75 (the same rapidity range of their spectrometer) to the
following expression: $Ed^3\sigma_{pp\rightarrow hX}/d^3p = A\,(1+p_T/p_0)^{-n}\,p_T/\sqrt{p_T^2+a}$, 
with $A$ = 244.5 [mb GeV$^{-2}c^{3}$], $p_0$ = 2.188 [GeV/$c$], $a$ = 0.0085 [GeV$^{2}/c^2$],
and $n$  = 15.37. STAR {\it preliminary} reference p+p spectrum 
~\cite{star_hipt_62} used a Hagedorn power-law form, $A\,[1+p_T/p_0]^{-n}$,
with parameters $A$ = 292.48 [mb GeV$^{-2}c^{3}$], $p_0$ = 1.75 [GeV/$c$], 
and $n$  = 13.23. Fig.~\ref{fig:pp_h_params_62GeV} compares the 3 parametrizations
to the existing ISR inclusive charged hadron cross-sections at 63 GeV. The 
3 parametrizations agree among each other (and the data) within $\pm$20\%. 
A direct measurement of the inclusive charged hadron production at large $p_T$ in p+p collisions 
in a dedicated run at $\sqrt{s}$ = 62.4 GeV at RHIC is mandatory, however,
if one wants to reduce the corresponding uncertainties propagated to the
Au+Au nuclear modification factors and constrain more quantitatively the model 
predictions for the parton energy loss excitation function (see Section~\ref{sec:R_AA_62.4}).

\clearpage

\subsection{p+p $\rightarrow$ $\pi^0 +X$ reference at $\sqrt{s}$ = 62.4 GeV}
\label{sec:pp_pi0_ref_62GeV}

\begin{figure}[htbp]
\begin{center}
\includegraphics[height=7.0cm]{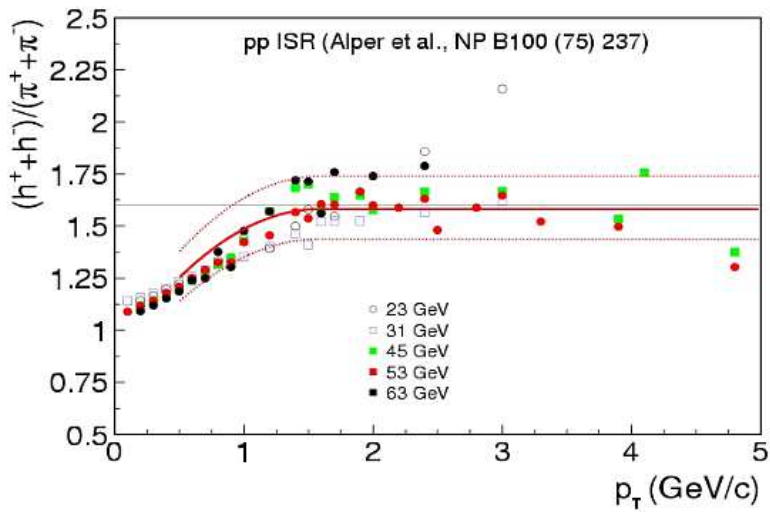}\hfill
\end{center}
\caption{Ratio of total charged hadron over pion spectra in p+p collisions at ISR 
energies ($\sqrt{s}$ = 21, 31, 44, 53, and 63 GeV~\protect\cite{alper75}).
The straight line is at $h^\pm/\pi$ = 1.6.}
\label{fig:hpi_ratio_isr}
\end{figure}

\begin{figure}[h]
\begin{center}
\includegraphics[height=10.cm]{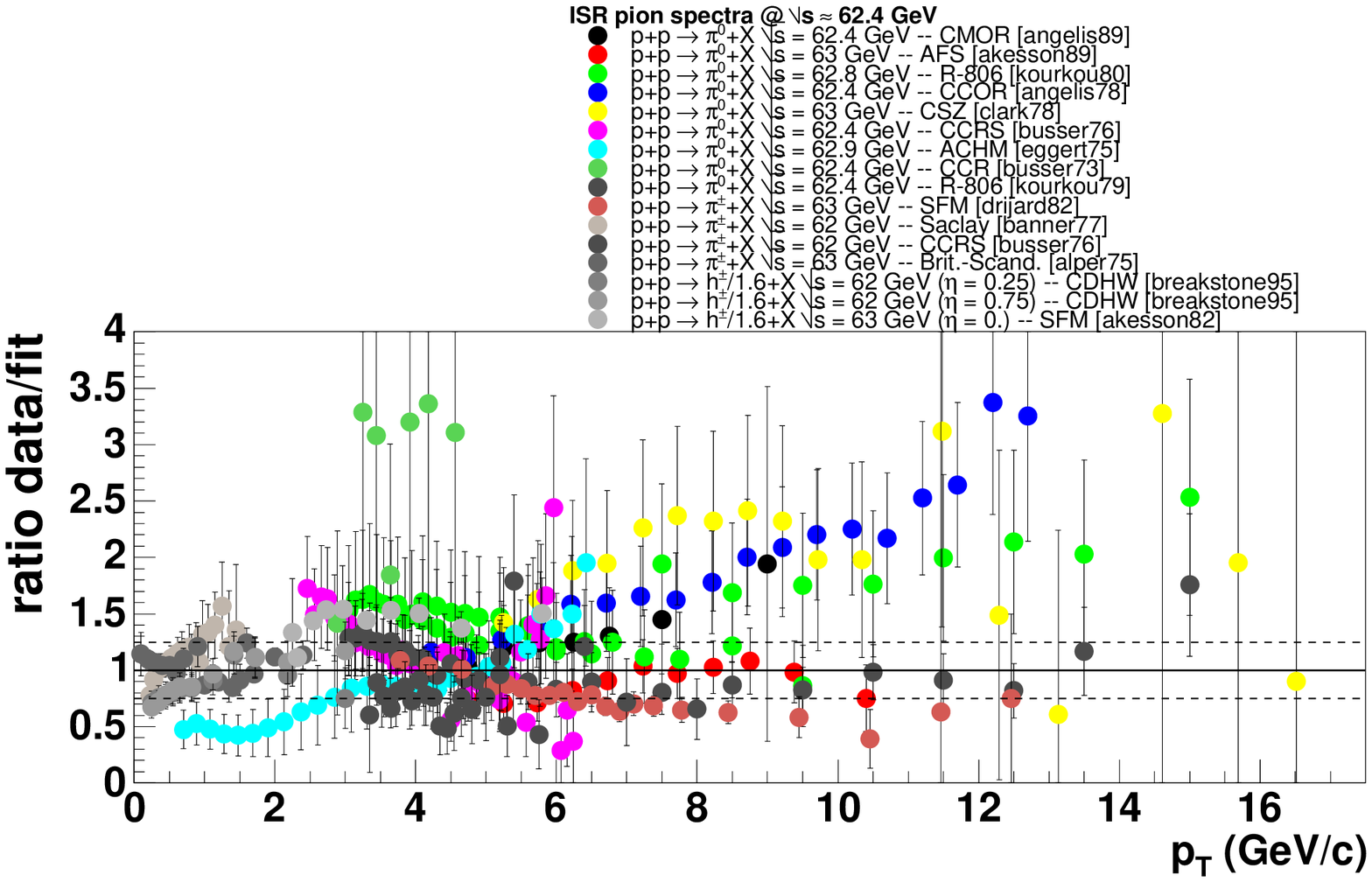}\hfill
\includegraphics
[height=6.7cm,clip=true]{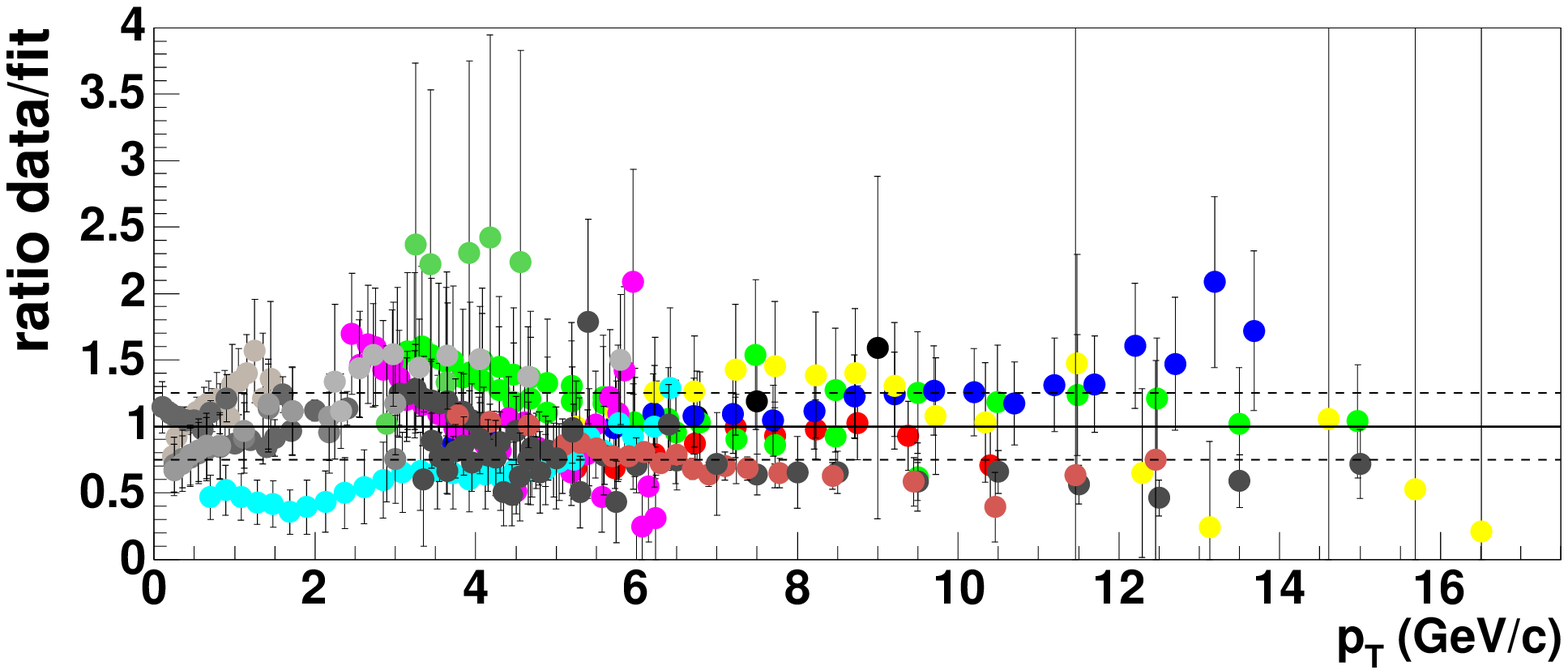}\hfill
\end{center}
\caption{Ratio of all experimental measurements reported in Table~\ref{tab:compilation}
over the final p+p $\rightarrow\pi^0$ parametrization (Eq.~(\ref{eq:final_fit})). 
The upper plot shows the data as originally published whereas the lower panel
shows the corrected data as described in the text. The error bars for each data set
include (in quadrature) the original point-to-point, systematic and normalization 
uncertainties.}
\label{fig:ratio_fit_all_data_sets}
\end{figure}

In order to obtain a benchmark p+p$\rightarrow\pi^0+X$ reference 
spectrum at $\sqrt{s}$ = 62.4 GeV we first collected (Table~\ref{tab:compilation}) 
all existing experimental $\pi^0$ (9 measurements~\cite{busser73,busser74,
eggert75,busser76,clark78,angelis78,kourkou79,kourkou80,angelis89,akesson89}), $\pi^\pm$ (4 
measurements~\cite{busser76,alper75,banner77,drijard82}) and $h^\pm$ (3 
measurements~\cite{akesson82,breakstone95}) at the highest ISR collider energies 
($\sqrt{s}$ = 62 -- 63 GeV) and added in quadrature the original systematic 
and normalization uncertainties to the point-to-point errors. 
We included in our compilation the averaged ($\pi^+ +\pi^-$)/2 
spectra\footnote{Isospin symmetry justifies the assumption: 
$\pi^0 \approx (\pi^++\pi^-)/2$.} as well as the inclusive charged hadron spectra 
divided by the measured $h^\pm/\pi$ = 1.6 $\pm$ 0.16 ratio (Fig.~\ref{fig:hpi_ratio_isr}) 
since this provided us with additional constraints for our global fit analysis
at relatively low $p_T$ where no neutral pion data is available. 
The corresponding data points (adding to a total of $\sim$300) were 
fitted to a common functional form. The ratio of the weighted average
fit over each data set is shown in the upper plot of 
Fig.~\ref{fig:ratio_fit_all_data_sets}. Large differences in the shape 
of the $p_T$ distributions and in the magnitude of the cross-sections
are evident which implies that many of the measurements are 
inconsistent among each other well beyond the originally quoted uncertainties.

Investigation of the original published results indicates, however, 
several relevant differences affecting the experimental measurements.
First, although at first sight discrepancies of order $\sim$1 GeV in
the center-of-mass energy should not dramatically modify the single $\pi^0$
spectra, it turns out that at very high $p_T$ the absolute
difference in the perturbative yields between $\sqrt{s}$ = 62 GeV and 63 GeV
can indeed be as large as $\sim$20\% (Fig.~\ref{fig:ratio_sqrts}). 
Secondly and most important, only one experiment (Akesson {\it et al.}~\cite{akesson89}) 
fully identified neutral pions via a standard invariant mass analysis of 
photon pairs. The rest of the experiments either did not separate the 
$\gamma\gamma$ decay of $\pi^0$ and $\eta$ and/or did not subtract 
the direct photon component from the experimentally measured 
``unresolved'' $\pi^0$ spectra. In order to do a meaningful comparison and 
average of all $\pi^0$ data sets, one needed therefore to subtract 
each one of these experimental ``contaminations'' from the published 
data tables. 

\clearpage
\subsubsection{Center-of-mass energy correction:}
\begin{figure}[htbp]
\begin{center}
\includegraphics[height=8.0cm]{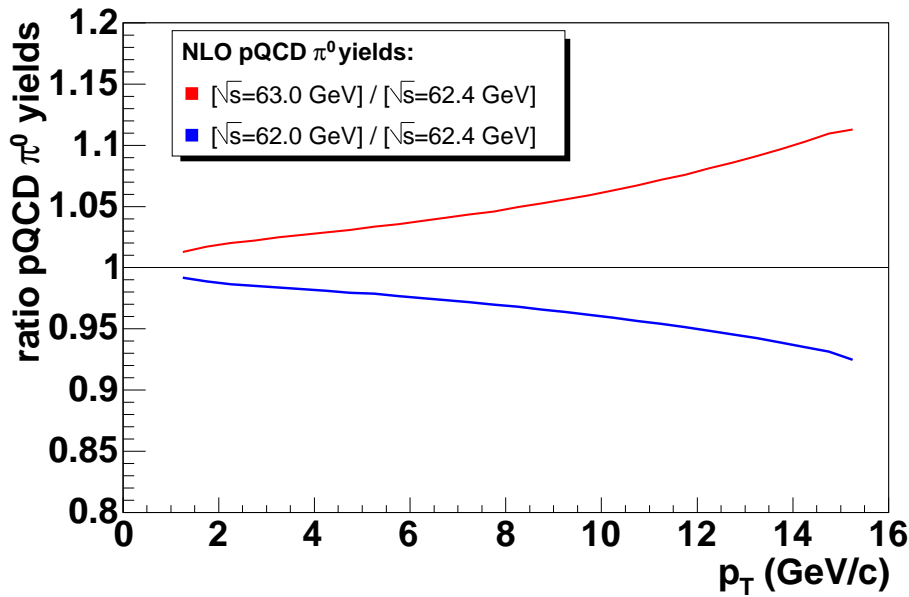}\hfill
\end{center}
\caption{Perturbative ratios of p+p $\rightarrow \pi^0$ yields vs. $p_T$ at two
different center-of-mass energies ($\sqrt{s}$ = 63 GeV, upper curve, and 62 GeV, 
lower curve) over the yields at $\sqrt{s}$ = 62.4 GeV, as given by NLO pQCD
calculations~\cite{vogelsang_pi0} (note that any uncertainties in the PDFs and/or FFs 
basically cancel out in the ratio).}
\label{fig:ratio_sqrts}
\end{figure}

CERN ISR was a collider with proton beams crossing (and colliding) with 
a non-null angle depending on the experimental runs and setups. 
This fact resulted in differences in the effective available
energy in the center-of-mass as large as $\sim$1 GeV
(from $\sqrt{s}$ = 62 to 63 GeV). RHIC Au+Au collisions were instead
performed at a fixed $\sqrt{s_{_{NN}}}$ = 62.4~GeV.
Fig.~\ref{fig:ratio_sqrts} shows the perturbative ratio of $\pi^0$ cross-sections
as a function of $p_T$ for p+p collisions at $\sqrt{s}$ = 62 and 63 GeV 
over those at $\sqrt{s}$ = 62.4 GeV as given by NLO calculations~\cite{vogelsang_pi0}. 
Whereas the effect on the yields is minimal ($\lesssim$5\%) for $p_T<$ 8 GeV/$c$,
the difference in yields monotonically increases with $p_T$ reaching a maximum
of $\sim$10\% at the highest measured $p_T$ values at ISR.
We corrected the yields of all data sets indicated in the last column 
of Table~\ref{tab:compilation} using a simple second-order polynomial 
$p_T$ fit of the computed pQCD ratio.

\subsubsection{Direct photon subtraction:}
\begin{figure}[htbp]
\begin{center}
\includegraphics[height=9.0cm]{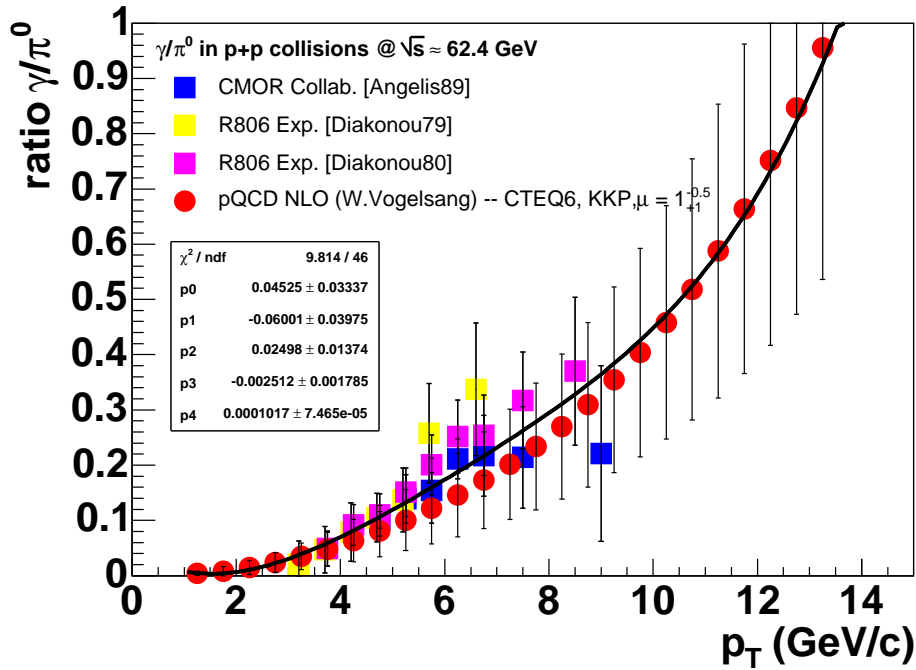}\hfill
\end{center}
\caption{Ratio of $\gamma/\pi^0$ cross-sections in p+p collisions 
at $\sqrt{s}\approx$ 62.4 GeV as a function of $p_T$ measured 
experimentally~\protect\cite{diakonou79,diakonou80,angelis89} (squares), 
and computed in NLO pQCD (circles, with error bars covering the range of 
theoretical scale uncertainties: $\mu$ = $p_T/2 - 2p_T$)~\protect\cite{vogelsang_pi0,vogelsang_gamma}. 
The black curve is a fit of the experimental and theoretical results to a 
common 4th order polynomial.}
\label{fig:gamma_pi0}
\end{figure}

Direct $\gamma$ were discovered in p+p at CERN-ISR in 1979~\cite{diakonou79}, 
therefore before this date any high-$p_T$ photon-like cluster 
detected in the electromagnetic calorimeters was ``identified'' as a neutral pion.
Figure~\ref{fig:gamma_pi0} shows the direct-$\gamma$/$\pi^0$ ratio as a function
of $p_T$ measured at $\sqrt{s}\approx$ 62.4 GeV by three 
ISR experiments (squares)~\cite{diakonou79,diakonou80,angelis89} compared to
the NLO pQCD ratio (circles) computed with CTEQ6 PDF and for three different 
(factorization-renormalization) scales (the theoretical points are centered 
at $\mu$ = $p_T$ and the ``errors'' correspond to $\mu$ = $p_T$/2 -- 2$p_T$)
~\cite{vogelsang_pi0,vogelsang_gamma}. The prompt photon ``contamination'' 
is marginal below $\sim$4 GeV/$c$, but it accounts for $\sim$1/3 of the $\pi^0$
yield at $p_T\sim$10 GeV/$c$, and equals it at $\sim$14 GeV/$c$. The agreement 
data--theory is good and allows us to extrapolate the ratio to  $p_T$ values 
higher than those measured in~\cite{diakonou79,diakonou80,angelis89}. 
The combined experimental and theoretical data points have been thus 
fitted to a 4th order polynomial (black curve) as a function of 
$p_T$, $R_{\gamma/\pi^0}(p_T)$, with parameters: $p_0$ = 4.55e-02, $p_1$ = -6.04e-02, 
$p_2$ = 2.51e-02, $p_3$ = -2.53e-03 and $p_3$ = 1.03e-04. 
The corrected $\pi^0$ yields, $Y_{\pi^0}(p_T)$, were obtained from the ``unresolved'' 
$\pi^0$ yields, $Y_{\pi^0+\gamma}(p_T)$, via: 
$Y_{\pi^0}=Y_{\pi^0+\gamma}\cdot R_{\gamma/\pi^0}^{-1}/(1+R_{\gamma/\pi^0}^{-1})$, 
for those data sets listed in the 11th column of Table~\ref{tab:compilation}.

\subsubsection{$\eta\rightarrow \gamma\gamma$ subtraction:}
\label{sec:eta_correction}

\begin{figure}[htbp]
\begin{center}
\includegraphics[height=7.9cm,width=15.0cm]{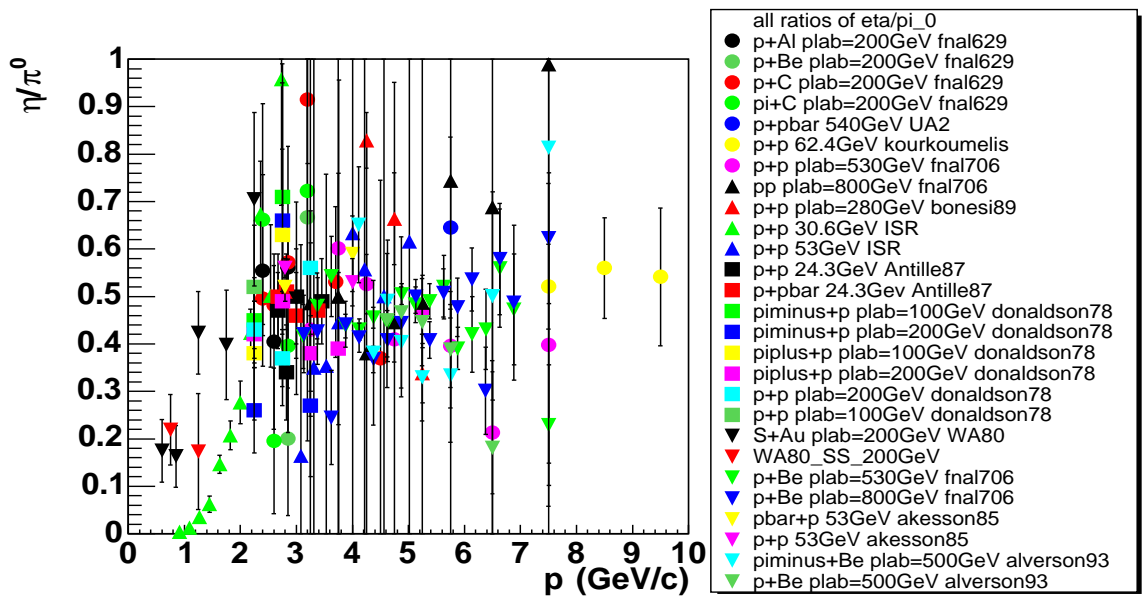}\hfill
\end{center}
\caption{Collected systematics of $\eta/\pi^0$ ratios in hadronic collisions. 
The average above $p_T$ = 1.5 GeV/$c$ is $R_{\eta/\pi^0}$ = 0.46.}
\label{fig:eta_pi0_ratio}
\end{figure}

Many of the ``unresolved'' $\pi^0$ measurements at ISR assumed
that all detected electromagnetic clusters at high $p_T$ were (merged)
photons from the $\pi^0$ decay and neglected any possible contribution 
from the 2-gamma decay channel of the $\eta$ meson (BR$_{\eta\rightarrow \gamma\gamma}$ = 0.394).
At high $p_T$ ($p_T>$ 1.5 GeV/$c$), however, the $\eta/\pi^0$  ratio is
$R_{\eta/\pi^0}$ = 0.46 (see the ``world'' compilation in Fig.~\ref{fig:eta_pi0_ratio}).
Those ``unresolved'' $\pi^0$ spectra listed in (the 12th column of) 
Table~\ref{tab:compilation} have been scaled {\it down} by a factor of 0.82 
above $p_T$ = 1.5 GeV/$c$ to take into account the 
(BR$_{\eta\rightarrow \gamma\gamma} \cdot R_{\eta/\pi^0} = 0.394\cdot 0.45$ = 0.18)
$\eta$ contamination [Note that such $\sim$18\% factor was mentioned 
in some of the original papers (e.g.~\cite{busser73,busser74}) 
but not actually subtracted from the tabulated results]. 

\subsubsection{Final parametrization:}

\begin{figure}[htbp]
\begin{center}
\includegraphics[height=10.0cm]{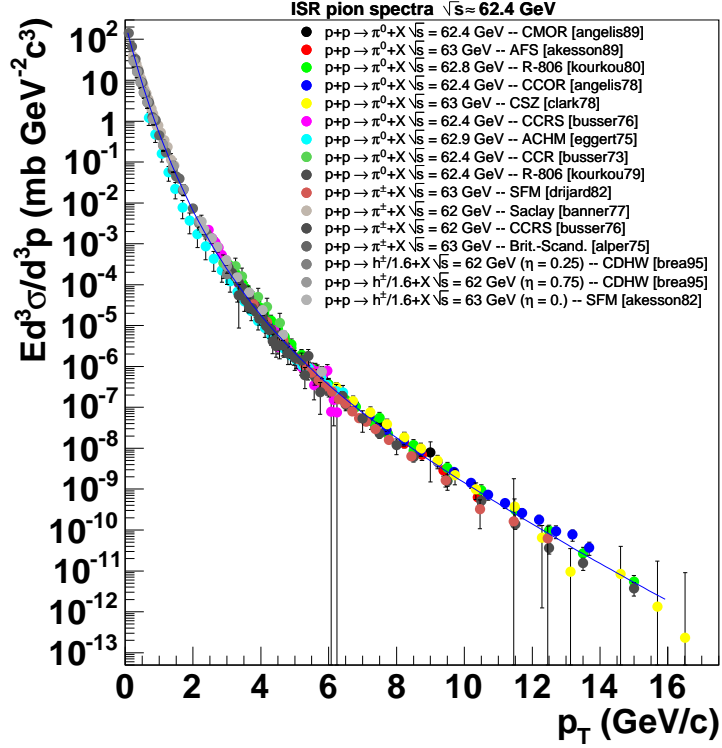}\hfill
\end{center}
\caption{Compilation of all pion transverse spectra measured in p+p
collisions at $\sqrt{s}\approx$ 62.4 GeV and fitted to a common
function, Eq.~(\ref{eq:final_fit}), with the parameters quoted in the text.}
\label{fig:final_fit_all_data_sets}
\end{figure}

\begin{figure}[htbp]
\begin{center}
\begin{minipage}[c]{.49\linewidth}
\includegraphics[height=10.0cm,width=9.0cm]{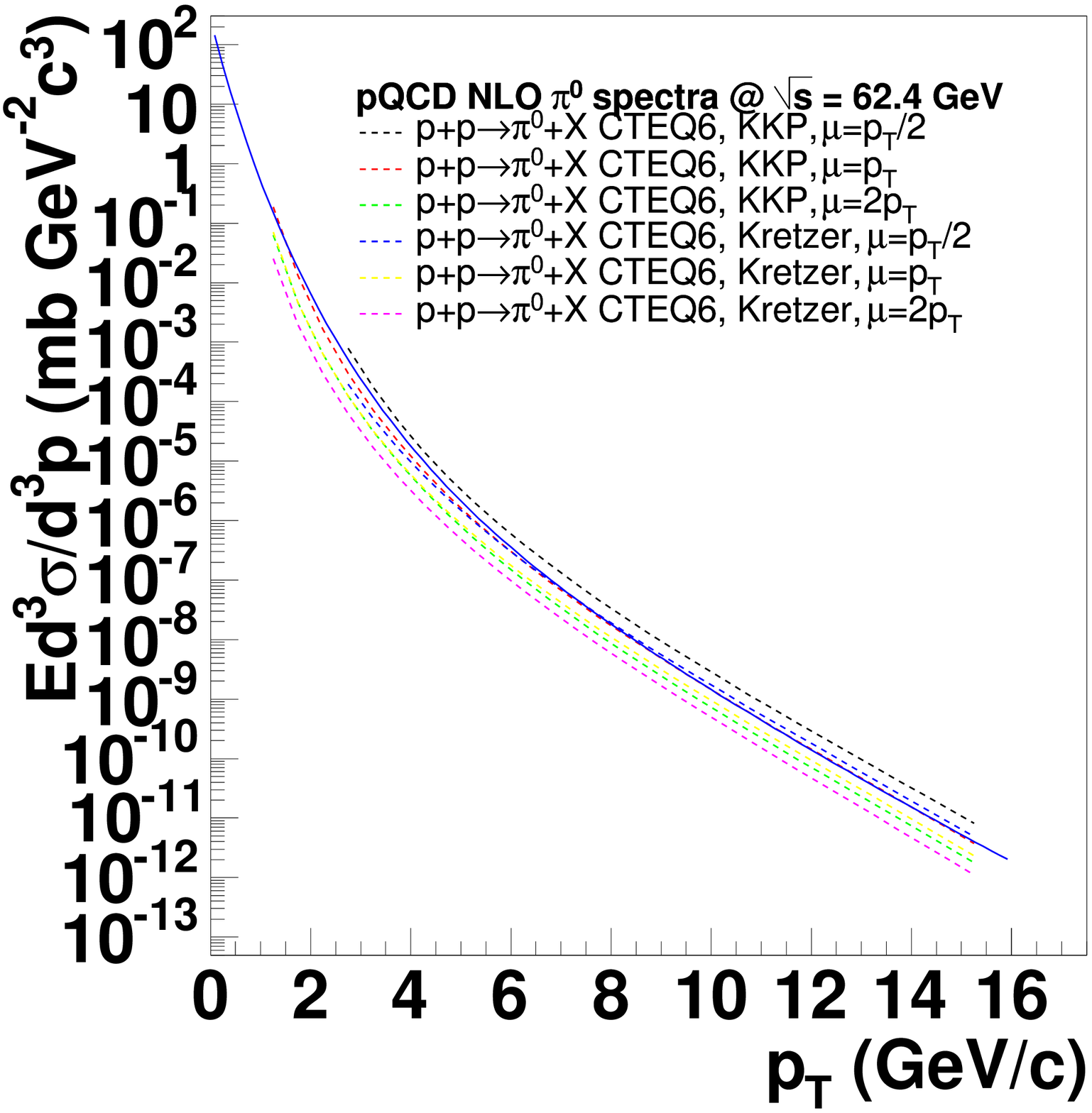}\hfill
\end{minipage} \hfill
\begin{minipage}[c]{.43\linewidth}
\includegraphics[height=6.0cm,width=7.50cm]{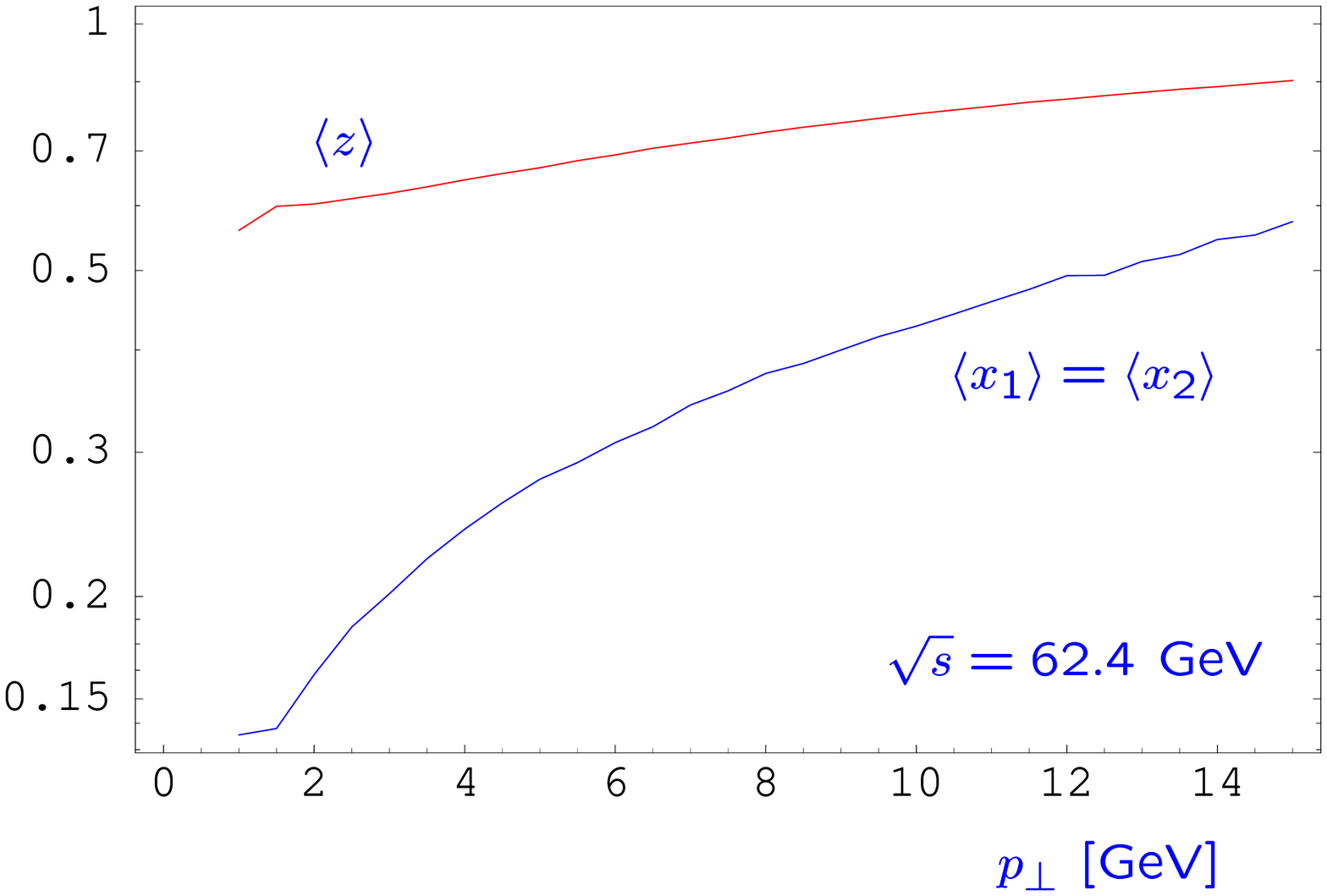}\hfill
\end{minipage} \hfill
\end{center}
\caption{Left: Comparison of the final empirical p+p $\rightarrow \pi^{0}+X$ 
parametrization at $\sqrt{s}$ = 62.4 GeV (solid curve)
to NLO pQCD calculations~\protect\cite{vogelsang_pi0} for 2 sets of FFs (KKP~\protect\cite{kkp} 
and Kretzer~\protect\cite{kretzer}) and 3 different scales $\mu$ = $p_T/2, p_T, 2p_T$.
Right: Scaling variables $\langle x_{1,2} \rangle$ (average parton fractional
momentum) and $\langle z \rangle$ (average momentum fraction of the parent parton
carried by the leading pion) for p+p $\rightarrow \pi^0$ ($\sqrt{s}$ = 62.4 GeV)
at mid-rapidity versus the $\pi^0$ momentum, computed in perturbative 
QCD~\protect\cite{kretzer0}.}
\label{fig:fit_vs_nlo}
\end{figure}

After correction of the experimental spectra as described in the previous
sections, there remained still a few outliers measurements that were
(partially or totally) inconsistent (beyond $\sim 1.5\sigma$) with the 
rest of the data (see lower panel of Fig.~\ref{fig:ratio_fit_all_data_sets}).
Three data sets: Drijard {\it et al.}~\cite{drijard82} (charged pions measured off
central rapidity, at 50$^o$), Eggert {\it et al.}~\cite{eggert75} and 
Busser {\it et al.}~\cite{busser73} were excluded of the final global fit analysis.
Fig.~\ref{fig:final_fit_all_data_sets} shows all the (corrected) 
pion data sets plotted as a function of $p_T$ compared to a common
(purely empirical) 5-parameter functional form:

\begin{equation}
Ed^3\sigma_{pp\rightarrow \pi X}/d^3p = A\,(e^{a\cdot p_T^2 + b \cdot p_T}+p_T/p_0)^{-n},
\label{eq:final_fit}
\end{equation}

\noindent
with parameters: $A$ = 265.1 [mb GeV$^{-2}c^{3}$], $a$ = -0.0129 [GeV$^{-1}c$], 
$b$ = 0.04975 [GeV/$c$], $p_0$ = 2.639 [GeV/$c$], and $n$ = 17.95.
Eq.~(\ref{eq:final_fit}) provides a very good reproduction\footnote{Mind 
that this is a purely empirical fit whose validity beyond the 
considered $p_T$ range is {\it not} guaranteed.} of the full spectral shape 
in the range $p_T$ = 0 -- 16 GeV/$c$. Above $p_T\approx$ 8 GeV/$c$ the data
(and the fit) have an exponential-like shape. This departure from the pure 
power-law behaviour expected for parton-parton scatterings is due to the fact 
that such large $p_T$ values are in a kinematical domain above 
$\langle x_T \rangle\approx$ 0.3 and $\langle z \rangle\approx$ 0.8 (see 
Fig.~\ref{fig:fit_vs_nlo}, right) where both the parton distribution functions 
(PDFs) and fragmentation functions (FFs) resp. start to decrease due to phase 
space boundaries (PDFs and FFs reach zero at the kinematical limit: 
$p_T\approx$ 30 GeV/$c$). Such a change in the $p_T$ shape is confirmed also 
by pQCD calculations (see Fig.~\ref{fig:fit_vs_nlo}).

\begin{figure}[htbp]
\begin{center}
\includegraphics[height=8.5cm,width=14.0cm]{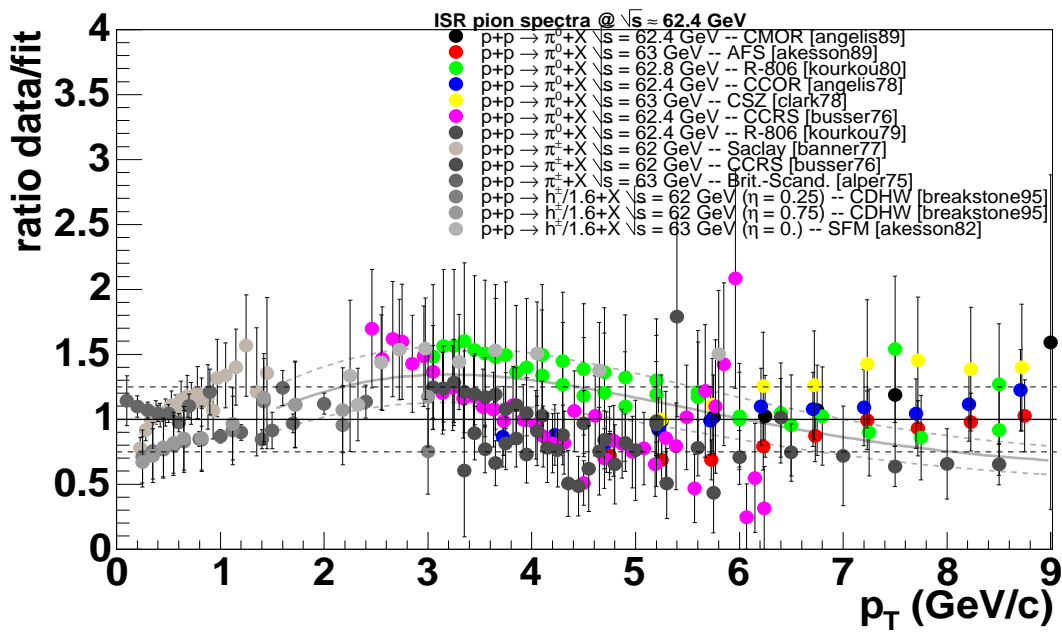}\hfill
\end{center}
\caption{Ratio of the selected and corrected $\pi^{0,\pm}$ and (scaled) $h^\pm$ measurements 
at CERN-ISR over the final $\pi^0$ parametrization (Eq.~(\ref{eq:final_fit}) 
with fit parameters reported in the text) as a function of $p_T$. 
The gray line is the charged-hadron reference spectrum (Section~\ref{sec:pp_h_ref_62GeV}),
divided by $h^\pm/\pi$ = 1.6 $\pm$0.16 (The errors of the $h^\pm/\pi$ ratio are shown as 
dashed gray lines. There is an additional systematic 25\% error in the $h^\pm$ 
reference not plotted).}
\label{fig:fit_vs_data_zoom}
\end{figure}

Figure~\ref{fig:fit_vs_data_zoom} shows the ratio of the (selected and 
corrected) experimental pion spectra over the final p+p $\rightarrow \pi^{0}+X$ 
parametrization in the same $p_T$ range ($p_T$ = 1 -- 8 GeV/$c$) covered 
by the PHENIX Au+Au $\rightarrow\pi^0+X$ data at RHIC. All p+p cross-sections 
are consistent with the final fit within their associated errors.
The dashed lines indicate the $\pm$25\% systematic uncertainty assigned 
to the reference. The gray thick line is the PHENIX charged-hadron 
reference (see Section~\ref{sec:pp_h_ref_62GeV}) divided by the expected 
$h^\pm/\pi^0$ = 1.6 $\pm$ 0.16 ratio at $p_T>$ 1.5 GeV/$c$ (see 
Fig.~\ref{fig:hpi_ratio_isr}). The $\pm$0.16 errors of the $h^\pm/\pi$ ratio 
are shown as dashed gray lines. The additional systematic 25\% error in the 
$h^\pm$ reference is not plotted.

\begin{figure}[htbp]
\begin{center}
\includegraphics[height=7.5cm,width=11.0cm]{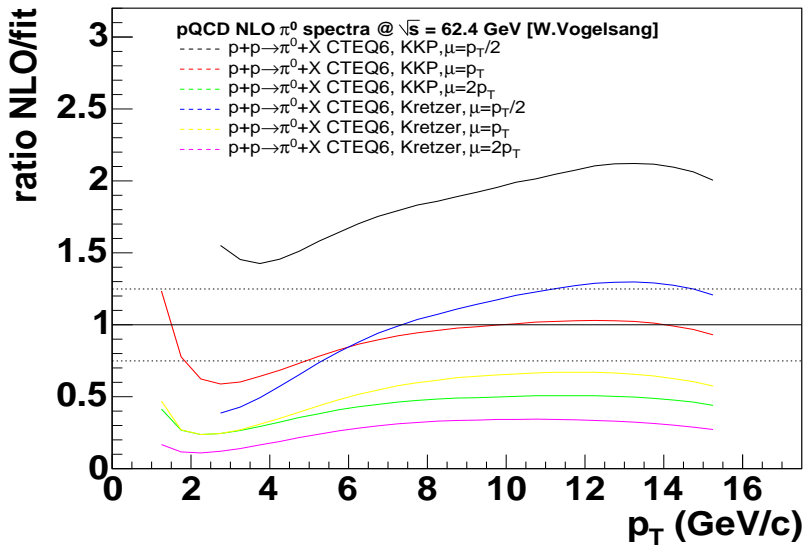}\hfill
\end{center}
\caption{Ratio of the NLO pQCD predictions for inclusive p+p $\rightarrow \pi^{0}+X$
production at $\sqrt{s}$ = 62.4 GeV (Fig.~\protect\ref{fig:fit_vs_nlo}) 
over the final empirical $\pi^0$ parametrization (Eq.~(\ref{eq:final_fit}
with the fit parameters reported in the text) as a function of $p_T$.}
\label{fig:fit_vs_nlo_ratio}
\end{figure}

\newpage
Figures~\ref{fig:fit_vs_nlo}-\ref{fig:fit_vs_nlo_ratio} show a comparison of the 
empirical $\pi^0$ parametrization to NLO pQCD predictions from W.~Vogelsang~\cite{vogelsang_pi0}
with fixed PDFs (CTEQ6), 2 different sets of fragmentation functions
(Kniehl-Kramer-P\"{o}tter KKP~\cite{kkp} and Kretzer~\cite{kretzer}) 
and 3 (factorization-renormalization-fragmentation) scales. 
The general shape and overall magnitude of the $p_T$ spectrum is well reproduced 
by the theoretical calculations (the fit is well contained within the theoretical
limits given by $\mu$ = $p_{T}/2 - 2p_T$). KKP FFs seem to reproduce better the 
magnitude of the cross-section, and the scale $\mu$ = $p_T$ provides the best
agreement with the data, especially above $p_T\approx$ 5 GeV/$c$. 
Below $p_T\approx$ 5 GeV/$c$ all NLO spectra tend to consistently underpredict 
the observed $\pi^0$ cross-section. Such a discrepancy is also observed 
in high $p_T$ hadro-production at lower $\sqrt{s}$~\cite{bourre} 
where soft-gluon resummation corrections~\cite{resumm} and additional 
non-perturbative effects (e.g. intrinsic $k_T$~\cite{e706_kt}) 
must be introduced to bring parton model analyses into accord with data.

\clearpage
\subsection{Nuclear modification factor at $\sqrt{s_{\mbox{\tiny{\it{NN}}}}}$ = 62.4 GeV:}
\label{sec:R_AA_62.4}

\begin{figure}[htbp]
\begin{center}
\epsfig{file=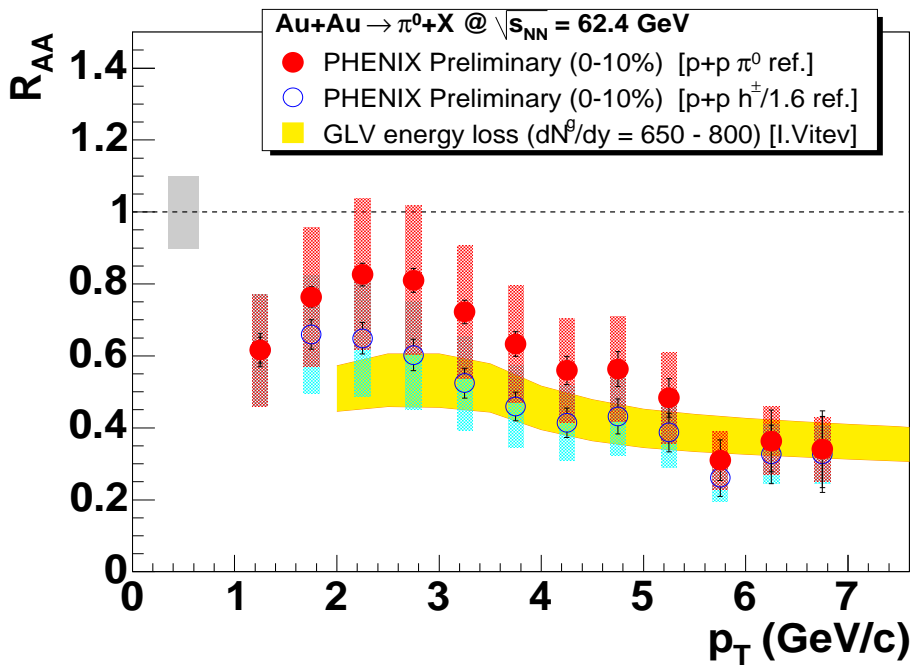,height=9.0cm,width=12.50cm}
\end{center}
\caption{Preliminary PHENIX nuclear modification factor, $R_{AA}(p_T)$, 
for $\pi^0$ measured in central Au+Au at 62.4 GeV~\protect\cite{phenix_hipt_62}
obtained using the p+p $\rightarrow \pi^{0}+X$ (red circles)
and p+p $\rightarrow h^{\pm}+X$ (open circles) references discussed in the
text, compared to theoretical predictions for parton energy loss in a 
dense medium with $dN^g/dy=$ 650 -- 800~\protect\cite{vitev04}.}
\label{fig:R_AA_pi0_62.4}
\end{figure}

Figure~\ref{fig:R_AA_pi0_62.4} shows the {\it preliminary} nuclear modification 
factor, Eq.~(\ref{eq:R_AA}), for high $p_T$ $\pi^0$ production in 0-10\% central Au+Au 
collisions at $\sqrt{s_{\mbox{\tiny{\it{NN}}}}}$ = 62.4 GeV~\cite{phenix_hipt_62} 
obtained using: (i) the p+p $\pi^0$ reference Eq.~(\ref{eq:final_fit}) (red circles), 
and (ii) the p+p $h^\pm$ reference (see Section~\ref{sec:pp_h_ref_62GeV}) divided by 
the expected $h^\pm/\pi^0$ = 1.6 ratio (open circles). [For complementary info, 
the average Glauber number of $NN$ collisions is $\langle N_{coll}\rangle$ = 845.4 $\pm$140
for a p+p inelastic cross-section at $\sqrt{s}$ = 62.4 GeV of $\sigma_{inel}$ = 35.6 $\pm$ 0.5 mb 
obtained from the weighted average values of the {\it total}~\cite{Eggert75,Amaldi76,Amaldi78,Baksay78,Carboni84,Ambrosio82,Amos85} 
($\sigma_{tot}$ = 43.37 $\pm$0.17 mb) and {\it elastic}~\cite{Baksay78,Amos85} 
($\sigma_{el}$ = 7.75 $\pm$0.10 mb) cross-sections measured at CERN-ISR]. 
Both $R_{AA}$ are compared to parton energy loss predictions for the quenching 
factor in a system with effective gluon densities $dN^g/dy=$ 650 -- 800 
(yellow band)~\cite{vitev04}. At intermediate
$p_T\approx$ 2 -- 5 GeV the use of one or the other parametrization results
in differences as large as $\sim$25\% in the amount of the suppression. 
Those divergences are indicative of the systematic uncertainty of the 
obtained p+p baseline references at  $\sqrt{s}$ = 62.4 GeV.
Clearly a dedicated RHIC proton-proton run at this collision energy would help to 
reduce these uncertainties and better constraint the theoretical predictions
of the excitation function of the high $p_T$ suppression~\cite{vitev04,adil_gyulassy04,wang04,dainese04}.

\clearpage

\section*{\underline{Case III}: p+p $\rightarrow$ $\gamma +X$
reference at $\sqrt{s}$ = 200 GeV}

\begin{figure}[htbp]
\begin{center}
\epsfig{file=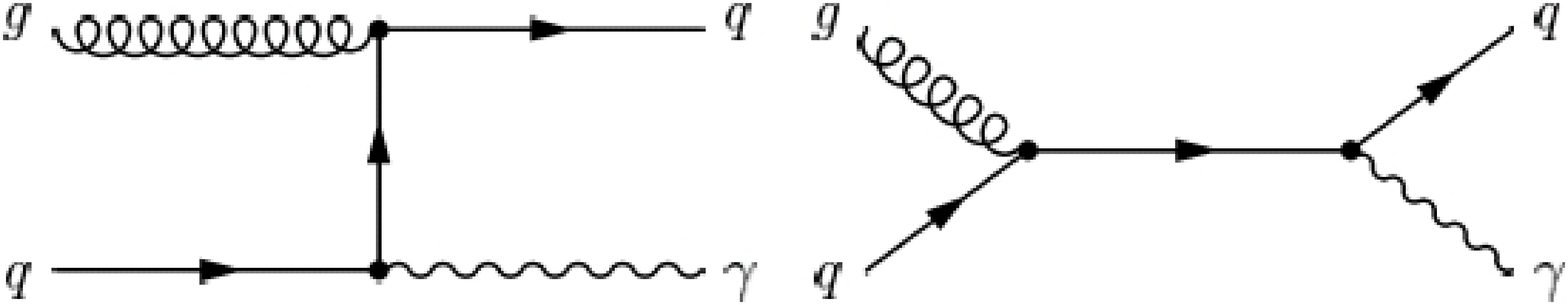,width=10.cm,height=2.0cm}
\vspace{0.5cm}
\epsfig{file=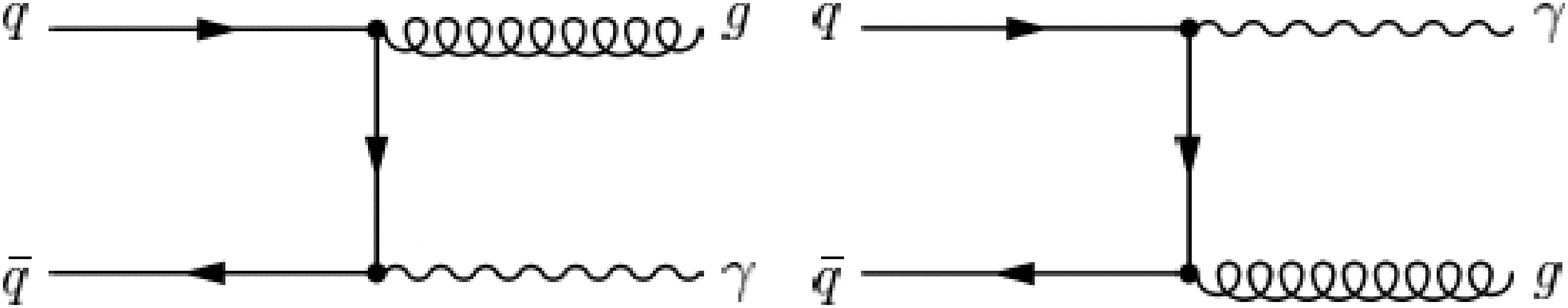,width=10.cm,height=2.0cm}
\end{center}
\caption{Compton and annihilation diagrams for direct photon production in parton-parton scatterings.}
\label{fig:photons_compton_annihilation}
\end{figure}

\begin{figure}[htbp]
\begin{center}
\epsfig{file=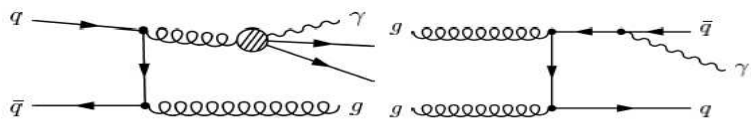,width=10.cm,height=2.0cm}
\end{center}
\caption{Fragmentation (or ``bremsstrahlung'') diagrams for direct photon production 
in parton-parton scatterings.}
\label{fig:photons_frag}
\end{figure}

Thermal (real or virtual) photons emitted in high-energy A+A reactions provide
direct information on the {\it thermodynamical} properties of the radiating 
underlying QCD matter and have long been considered privileged signatures 
of QGP formation~\cite{phot_reps}.
Direct photons, defined as real photons not originating from the decay of 
final hadrons, are emitted at various stages of a A+A reaction. 
Three qualitatively different mechanisms are usually considered: 
(i) prompt $\gamma$ (``pre-equilibrium'' or ``pQCD'') emission from perturbative 
parton-parton scatterings in the first tenths of fm/$c$ of the reaction, 
and (ii) subsequent emission from the thermalized partonic (QGP) and 
(iii) hadronic (hadron gas, HG) phases of the reaction. 
The partonic diagrams contributing in leading-order to photon production 
are $qg$-Compton and $q\bar{q}$-annihilation (Fig.~\ref{fig:photons_compton_annihilation}) 
and collinear $q,g$ fragmentation (Fig.~\ref{fig:photons_frag}).

\begin{figure}[htbp]
\begin{center}
\epsfig{file=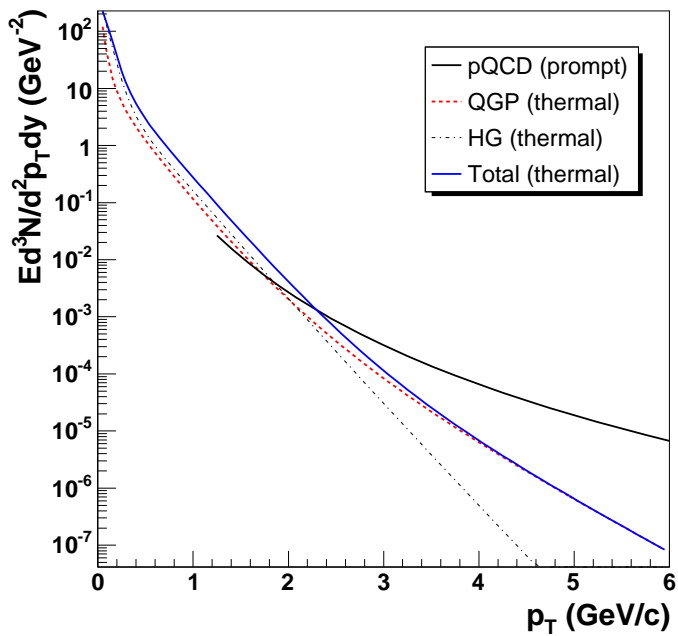,height=8.5cm}
\end{center}
\caption{Expected thermal and prompt photon spectrum for central Au+Au reactions
at $\sqrt{s_{\mbox{\tiny{\it{NN}}}}}$ = 200 GeV as given by a hydrodynamical
model calculation~\protect\cite{dente_peressou} complemented with pQCD yields
for the prompt $\gamma$~\protect\cite{vogelsang_gamma}.}
\label{fig:thermal_photon_spec}
\end{figure}

The expected photon spectrum from Au+Au reactions at $\sqrt{s_{\mbox{\tiny{\it{NN}}}}}$ = 200 GeV,
can thus be obtained theoretically by combining: (i) NLO pQCD calculations for the 
primordial (hard) production (perturbative p+p yields~\cite{vogelsang_gamma} scaled by 
the nuclear overlap function $T_{AA}$), plus (ii) hydrodynamical calculations of the 
space-time evolution of the reaction~\cite{rasanen_sps_rhic,turbide,dente_peressou} 
complemented with modern parametrizations of the QGP~\cite{arnold} and HG~\cite{turbide} 
photon emission rates. Such calculations indicate that QGP thermal emission 
should be visible in a window of the inclusive direct photon spectrum 
between $p_T\approx$ 1 -- 3 GeV/$c$ (Fig.~\ref{fig:thermal_photon_spec}).

\begin{figure}[htbp]
\begin{center}
\epsfig{file=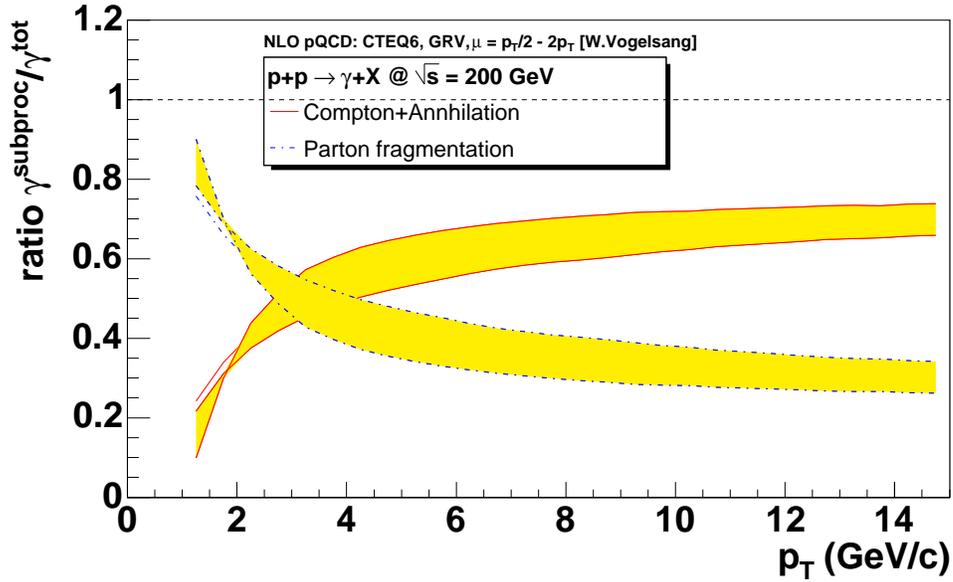,height=8.cm}
\end{center}
\caption{Relative contribution of different subprocesses of direct $\gamma$ production 
versus $p_T$ in p+p collisions at $\sqrt{s}$ = 200 GeV according to NLO pQCD~\protect\cite{vogelsang_gamma}:
Compton+Annihilation diagrams (upper band) and fragmentation diagrams (lower band).}
\label{fig:photons_frag_vs_dir}
\end{figure}

\begin{figure}[htbp]
\begin{center}
\epsfig{file=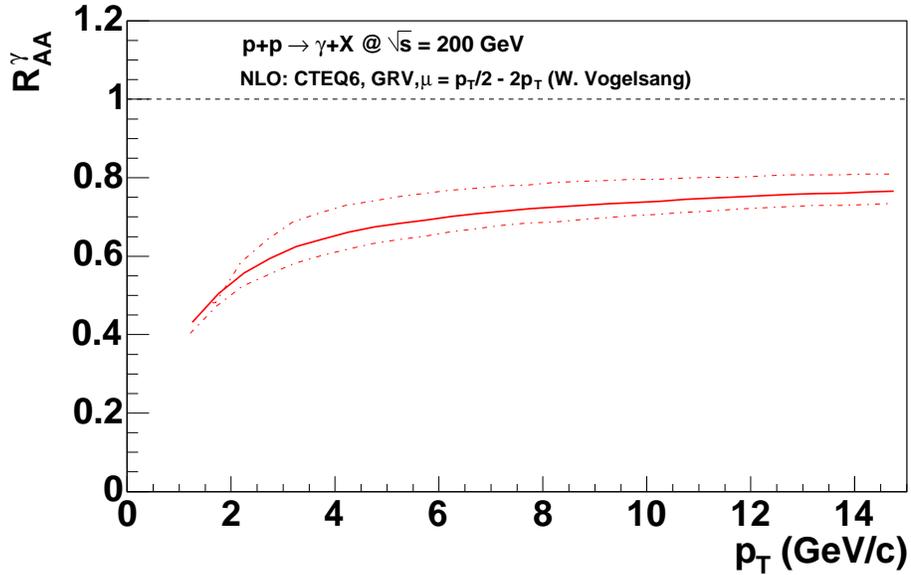,height=8.0cm}
\end{center}
\caption{Nuclear modification factor, $R_{AA}^{\gamma}(p_T)$, for direct $\gamma$ 
in central Au+Au collisions at $\sqrt{s_{\mbox{\tiny{\it{NN}}}}}$ = 200 GeV 
assuming the same quenching factor for the jet-fragmentation photon component 
as observed for high $p_T$ hadrons ($R_{AA}\approx$ 0.25).}
\label{fig:R_AA_photons_RHIC}
\end{figure}

However, unfortunately, this range of transverse momenta potentially presents 
difficulties for the extraction of the thermal component. Indeed,
perturbative QCD calculations indicate (Fig.~\ref{fig:photons_frag_vs_dir})
that below $p_T\approx$ 3 GeV/$c$, prompt photons are mainly produced via the 
parton bremsstrahlung mechanism (``anomalous'' component) [At increasingly high $p_T$, 
prompt photon production is dominated by the purely perturbative production 
mechanisms (Compton and annihilation, Fig.~\ref{fig:photons_compton_annihilation}) 
but still $\sim$1/4 of the total $\gamma$'s appear to come from jet fragmentation 
(diagrams depicted in Fig.~\ref{fig:photons_frag}) in these calculations]. Such a component
should be depleted in central Au+Au at $\sqrt{s_{\mbox{\tiny{\it{NN}}}}}$ = 200 GeV
due to the same final-state QCD medium effects that result in the observed factor 
of $\sim$4--5 suppression of high $p_T$ hadroproduction. 
To get a handle on the possible effect of parton energy loss in the total prompt 
photon spectrum we plot in Fig.~\ref{fig:R_AA_photons_RHIC} the ``photon nuclear
modification factor'', $R_{AA}^{\gamma}(p_T)$, for central Au+Au estimated simply assuming 
that the suppression factor for the $\gamma$-fragmentation component is the same as 
that observed for high $p_T$ hadrons, i.e. 
$R_{AA}^{\gamma_{frag}}=R_{AA}^{\mbox{\tiny{high $p_T$ $\pi^0$}}}\approx$ 0.25.
We determine $R_{AA}^{\gamma}$ = $[Ed\sigma_{\gamma_{tot}}/dp - 0.75\,Ed\sigma_{\gamma_{frag}}/dp]/[Ed\sigma_{\gamma_{tot}}/dp]$
with the same NLO yields~\cite{vogelsang_gamma} used in Figure~\ref{fig:photons_frag_vs_dir}.

Figure~\ref{fig:R_AA_photons_RHIC} indicates that one should expect a moderate
$\sim$30\% effective suppression of the total inclusive high $p_T$ photon 
spectra due to parton energy losses in the dense medium\footnote{Note 
that in this estimate we have not considered partially counteracting (small) 
effects such as ``Cronin enhancement''~\cite{dumitru} and Au PDF shadowing 
effects~\cite{eks98}.}. This estimate is consistent with more involved 
calculations of the same effect~\cite{jeon02,arleo04}. The suppression of the
jet bremsstrahlung component could, therefore, partially ``mask'' the enhancement 
due to thermal photon emission in the range $p_T$ = 1 -- 3 GeV/$c$. Since the
jet-fragmentation $\gamma$ component cannot be experimentally discarded
via the standard ``isolation'' method due to the large soft background in Au+Au 
collisions, the only way to disentangle experimentally the counterbalancing 
effects of the thermal and quenched prompt $\gamma$ in Au+Au requires a detailed
analysis of the p+p reference:

\begin{enumerate}
\item Measurement of {\it isolated} photons in p+p, $N_{pp\rightarrow\gamma_{isolated}}$,
down to $p_T\approx$ 1 GeV/$c$ with uncertainties $\lesssim$15\% provides a handle on 
the actual Compton+Annihilation reference production.
\item Measurement of {\it total} inclusive photons in p+p, $N_{pp\rightarrow\gamma_{total}}$, 
down to $p_T\approx$ 1 GeV/$c$ with uncertainties $\lesssim$15\% provides a handle 
(since $N_{pp\rightarrow\gamma_{tot}} = N_{pp\rightarrow\gamma_{isol}}+N_{pp\rightarrow\gamma_{fragm}}$) 
on the actual fragmentation $\gamma$ reference production.
\item Measurement of {\it total} inclusive $\gamma$ production in Au+Au, 
$N_{AuAu\rightarrow\gamma_{total}}$, down to $p_T\approx$ 1 GeV/$c$ 
with uncertainties $\lesssim$15\%.
\item The {\it upper} limit on {\it thermal} production for a given Au+Au centrality 
(with nuclear overlap function $T_{AA}$) is given by:
$N_{AuAu\rightarrow\gamma_{thermal}}^{max} = N_{AuAu\rightarrow\gamma_{total}} - T_{AA}\cdot N_{pp\rightarrow\gamma_{isolated}}$.
\item The {\it lower} limit on {\it thermal} production for a given Au+Au centrality 
(with nuclear overlap function $T_{AA}$) is given by:
$N_{AuAu\rightarrow\gamma_{thermal}}^{min} = N_{AuAu\rightarrow\gamma_{total}} - T_{AA}\cdot N_{pp\rightarrow\gamma_{total}}$.
\end{enumerate}

\begin{figure}[htbp]
\begin{center}
\begin{minipage}[c]{.44\linewidth}
\includegraphics[height=10.cm,width=8.0cm]{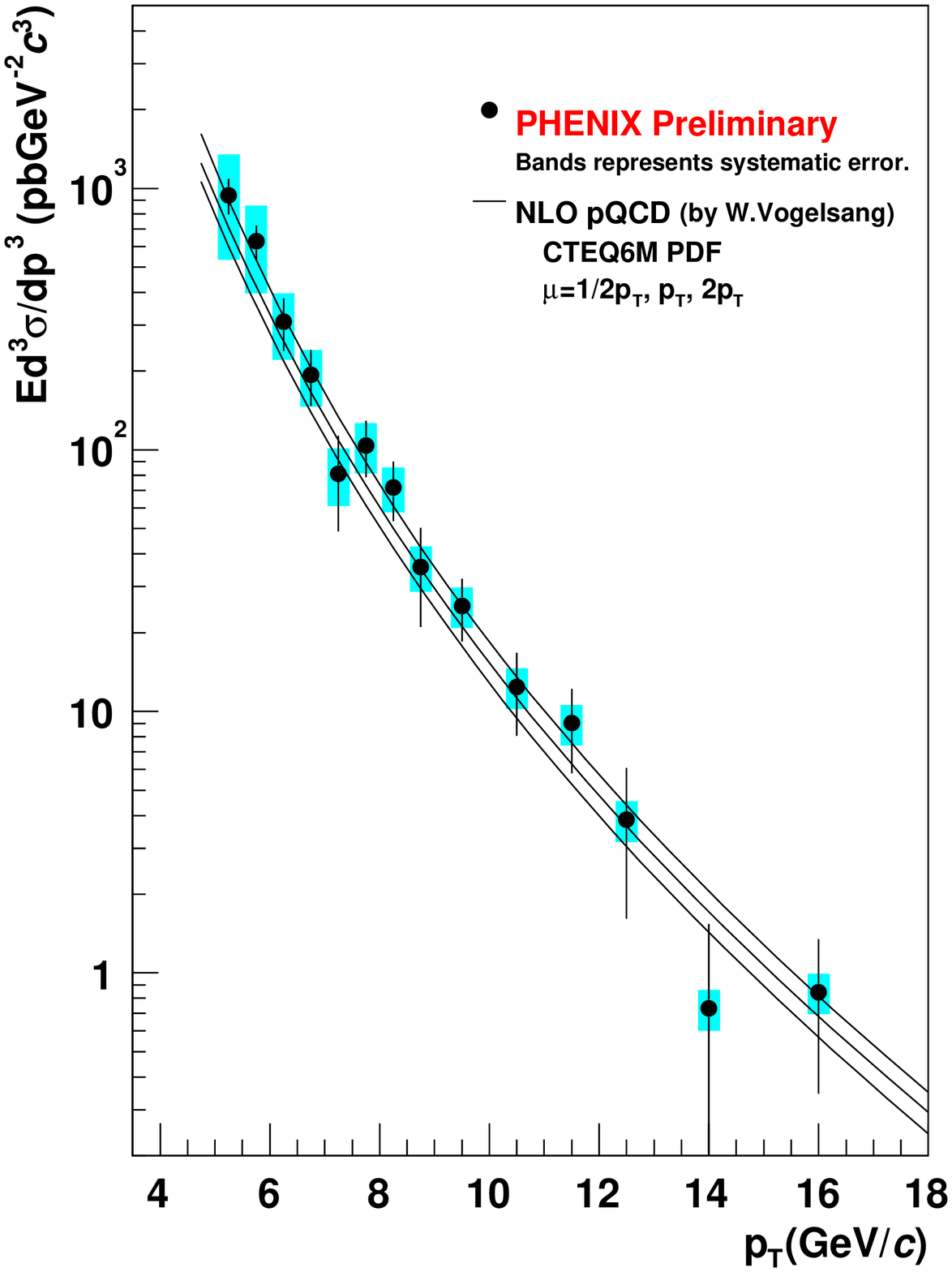}\hfill
\end{minipage} \hfill
\begin{minipage}[c]{.53\linewidth}
\includegraphics[height=7.50cm,width=11.0cm]{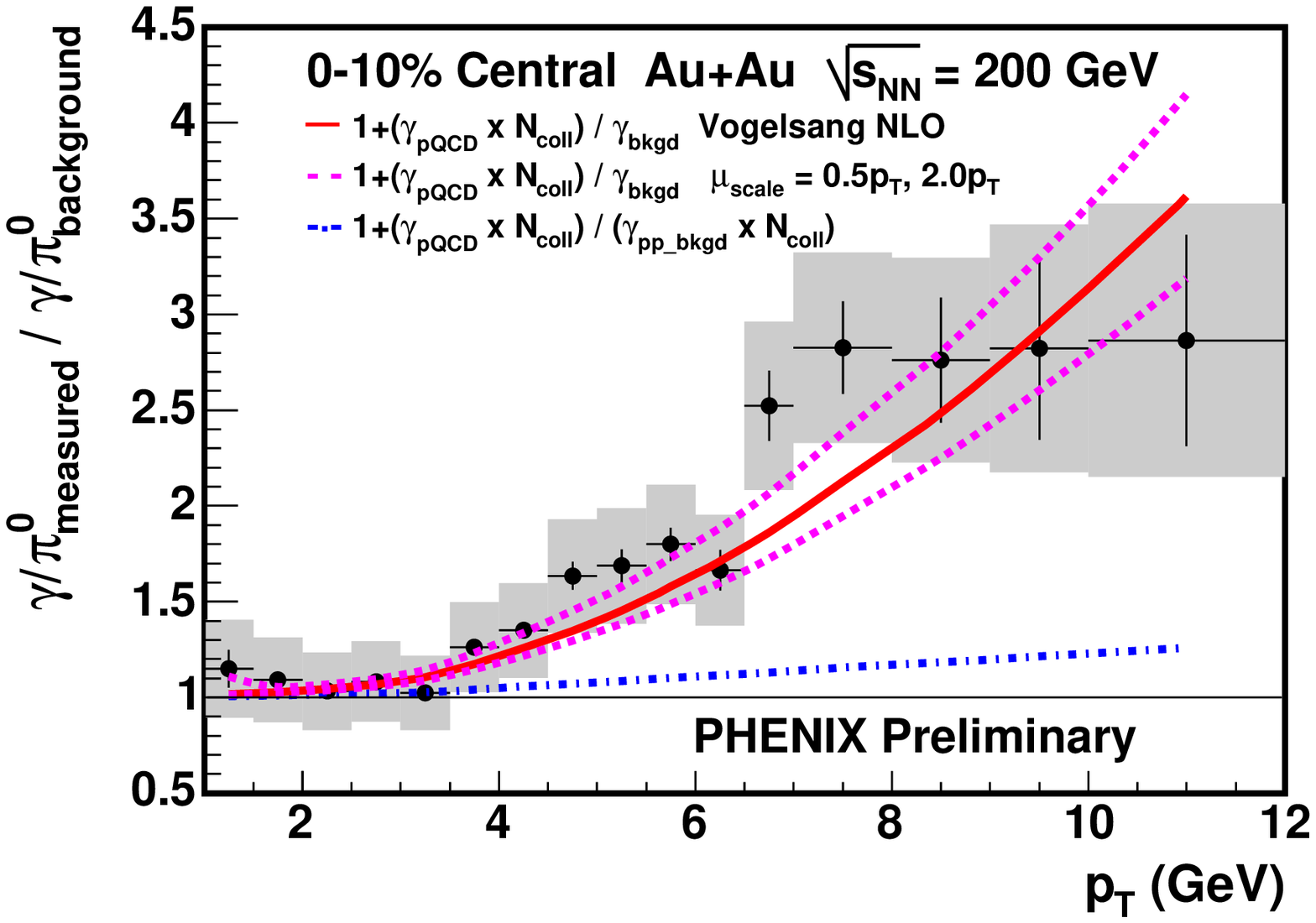}\hfill
\end{minipage} \hfill
\end{center}
\caption{Left: Preliminary p+p $\rightarrow \gamma+X$ measured by PHENIX at $\sqrt{s}$ = 200 GeV
~\protect\cite{okada04}. Right: Preliminary direct-$\gamma$ excess measured by PHENIX at 
$\sqrt{s_{\mbox{\tiny{\it{NN}}}}}$ = 200 GeV~\protect\cite{justin_qm04}.
Both results are compared to NLO pQCD calculations~\protect\cite{vogelsang_gamma}.}
\label{fig:pp_gamma_preliminary_phenix}
\end{figure}

Figure~\ref{fig:pp_gamma_preliminary_phenix} shows preliminary
measurements by PHENIX of the {\it total}~prompt photon production 
in p+p~\cite{okada04} [$N_{pp\rightarrow\gamma_{total}}$ in item (ii)] 
and in Au+Au~\cite{justin_qm04} [$N_{AuAu\rightarrow\gamma_{total}}$ 
in item (iii)] collisions at RHIC obtained by statistically subtracting 
the hadron decay-photon contributions ($\pi^0$, $\eta$, ...) 
from the total measured $\gamma$ spectrum. Within uncertainties, 
both results are consistent with the perturbative QCD 
expectations~\cite{vogelsang_gamma}. A separation of the p+p fragmentation-$\gamma$ 
component is under-way too~\cite{okada04}. The main issue in order to
experimentally resolve a possible thermal component in Au+Au is to measure 
with small uncertainties the prompt photon production below $p_{T}\approx$ 3 GeV/$c$ 
(where the background of decay photons and (anti)baryon contaminations is more 
significant) in both colliding systems. Complementary methods to the statistical 
subtraction one (e.g. the measurement via $\gamma$ ``conversion''), that are 
efficient for direct photon identification in the interesting range 
$p_{T}\approx$ 0.5 -- 2 GeV/$c$, are being currently pursued.

\section*{Summary}

Three different cases of hard production in proton-proton collisions
at three different center-of-mass energies, $\sqrt{s}\approx$ 20, 62 and 200 GeV,
have been discussed as benchmarks for the investigation of
QGP signals in nucleus-nucleus collisions. High $p_T$ neutral
pion production in central A+A collisions at CERN-SPS (
$\sqrt{s_{\mbox{\tiny{\it{NN}}}}}\approx$ 20 GeV) has been found to be 
(slightly) suppressed compared to a parametrization of the $\pi^0$ 
cross-sections in p+p collisions in free space. Such a result, possibly 
indicative of jet quenching effects at these energies, should be confirmed 
by a {\it direct} experimental measurement of high $p_T$ p+p 
hadroproduction at $\sqrt{s}\approx$ 20 GeV. Secondly, high $p_T$ 
$\pi^0$ and $h^\pm$ reference spectra at $\sqrt{s}$ = 62.4 GeV
have been constructed based on the weighted averaged of {\it revised} 
experimental data from CERN-ISR. Hard hadro-production is seen to be
suppressed by up to a factor of $\sim$3 in central Au+Au collisions at RHIC
at these collision energies. However, the current p+p references have 
$p_T$ dependent uncertainties of order $\sim$25\% precluding a detailed 
quantitative study of the excitation function of high $p_T$ nucleus-nucleus 
suppression between SPS and RHIC energies. A dedicated RHIC p+p run at 
$\sqrt{s}$ = 62.4 GeV would be needed to reduce such uncertainties
and better constraint the theoretical models of parton energy loss
in dense QCD matter. Finally, we have studied the possible effects
of parton energy loss in the direct photon contribution of jet-fragmentation
origin in Au+Au collisions at $\sqrt{s_{\mbox{\tiny{\it{NN}}}}}$ = 200 GeV. 
Estimates based on NLO pQCD indicate that the nuclear modification factor
can be reduced by $\sim$30\% due to these effects, partially hiding
the expected thermal photon emission from a radiating QGP.
We have discussed how to get a handle on a possible thermal radiation
Au+Au signal by detailed measurements of the isolated and non-isolated
direct photon spectra in baseline p+p collisions at $\sqrt{s}$ = 200 GeV.


\section*{Acknowledgments}
I would like to thank Werner~Vogelsang and Stefan~Kretzer for useful discussions 
and especially for providing with different pQCD calculations confronted
with the experimental data in this paper. Discussions with Angelika 
Drees on the difficulties of running RHIC in the proton-proton
mode at low energies are also acknowledged.



\section*{References}
\begin{thebibliography}{100} 

\def\IJMPA{{Int. J. Mod. Phys.}~{\bf A}}
\def\EPJ{{Eur. Phys. J.}~{\bf C}}
\def\JPG{{J. Phys}~{\bf G}}
\def\JHEP{{J. High Energy Phys.}~}
\def\NCA{Nuovo Cimento~}
\def\NIM{Nucl. Instrum. Methods~}
\def\NIMA{{Nucl. Instrum. Methods}~{\bf A}}
\def\NPA{{Nucl. Phys.}~{\bf A}}
\def\NPB{{Nucl. Phys.}~{\bf B}}
\def\PLB{{Phys. Lett.}~{\bf B}}
\def\PLC{Phys. Repts.\ }
\def\PRL{Phys. Rev. Lett.\ }
\def\PRD{{Phys. Rev.}~{\bf D}}
\def\PRC{{Phys. Rev.}~{\bf C}}
\def\ZPC{{Z. Phys.}~{\bf C}}

\bibitem{latt}See e.g. Karsch F 2002 {\it Lect. Notes Phys.} {\bf 583} 209 
\bibitem{factor}Collins J C, Soper D E and Sterman G 1985 {\it Nucl. Phys.} B {\bf 261} 104 
\bibitem{dde_qm04}d'Enterria D 2004 J.\ Phys.\ G {\bf 30}, S767, {\it Preprint}  nucl-ex/0404018 
\bibitem{na50_drellyan}NA50 Collaboration Bordalo P {\it et al.} 2003 Pramana {\bf 60} 817 
\bibitem{justin_qm04}PHENIX Collaboration Frantz J 2004 J.\ Phys.\ G {\bf 30}, S1003, {\it Preprint} nucl-ex/0404006
\bibitem{phnx_AuAucharm}PHENIX Collaboration Adler S S 2004 {\it et al.}, {\it Preprint} nucl-ex/0409028.
\bibitem{jet_quench_review}See review M.~Gyulassy, I.~Vitev, X.N.~Wang and B.W.~Zhang in Hwa R C (ed.) {\it et al.}: 
``Quark Gluon Plasma Vol. 3'', World Scientific, {\it Preprint}  nucl-th/0302077 
\bibitem{phenix_hipt_130}PHENIX Collaboration Adcox K {\it et al.} 2002 \PRL {\bf 88} 022301, {\it Preprint} nucl-ex/0109003
\bibitem{star_hipt_130}STAR Collaboration Adler C {\it et al.} 2002 \PRL {\bf 89} 202301, {\it Preprint} nucl-ex/0206011
\bibitem{phenix_hipt_pi0_200}PHENIX Collaboration Adler S S {\it et al.} 2003 \PRL {\bf 91} 072301, {\it Preprint} nucl-ex/0304022
\bibitem{star_hipt_200}STAR Collaboration Adams J {\it et al.} 2003 \PRL {\bf 91} 172302, {\it Preprint} nucl-ex/0305015
\bibitem{phobos_hipt_200}PHOBOS Collaboration Back B B {\it et al.} 2004 \PLB {\bf 578} 297, {\it Preprint} nucl-ex/0302015
\bibitem{brahms_hipt_200}BRAHMS Collaboration Arsene I {\it et al.} 2003 \PRL {\bf 91} 072305, {\it Preprint} nucl-ex/0307003
\bibitem{phenix_pp_pi0_200}PHENIX Collaboration Adler S S {\it et al.} 2003 {\it Phys. Rev. Lett.} {\bf 91} 241803, 
{\it Preprint} hep-ex/0304038 
\bibitem{wa98}WA98 Collaboration Aggarwal M M {\it et al.} 2002 {\it Eur. Phys. J. C} {\bf 23} 225 
WA98 collaboration Aggarwal M M {\it et al.} 1998 \PRL {\bf 81} 4087; [Erratum-ibid.\ {\bf 84} (2000) 578] 
\bibitem{wang_sps}Wang X N 1998 \PRL {\bf 81} 2655 
\bibitem{wang_syst}Wang X N 2000 \PRC {\bf 61} 064910; 
Wang E and Wang X N 2001 \PRC {\bf 64} 034901 
\bibitem{cronin}Cronin J W {\it et al.} 1975 {\it Phys. Rev.} D{\bf 11} 3105 
\bibitem{antrea}Antreasyan D {\it et al.} 1979 {\it Phys. Rev.} D {\bf 19} 764 
\bibitem{straub}Straub P B {\it et al.} 1992 \PRL {\bf 68} 452 
\bibitem{isr_alpha2_pi0}BCMOR Collaboration Angelis A L S {\it et al.} 1987 {\it Phys. Lett.} B 185 213 
\bibitem{dde_hipt_SPS}d'Enterria D 2004 Phys.\ Lett.\ B {\bf 596}, 32, {\it Preprint} nucl-ex/0403055 
\bibitem{hagedorn}Hagedorn R 1984 Riv.\ Nuovo Cim.\  {\bf 6N10} 1 
\bibitem{beier}Beier E W {\it et al.} 1978 Phys.\ Rev.\ D {\bf 18} 2235 
\bibitem{carey}Carey D C {\it et al.} 1976 Phys.\ Rev.\ D {\bf 14} 1196 
\bibitem{donaldson}Donaldson G {\it et al.} 1976 \PRL {\bf 36} 1110 
\bibitem{adams}FNAL E704 Collaboration Adams D L {\it et al.} 1996 Phys.\ Rev.\ D {\bf 53} 4747 
\bibitem{kretzer0}Kretzer S 2004 hep-ph/0410219, and private communication
\bibitem{blatt}Blattnig S R {\it et al.} 2000 Phys.\ Rev.\ D {\bf 62} 094030 
\bibitem{ceres}CERES/NA45 Collaboration Wurm J P and Bielcikova J 2004, {\it Preprint} nucl-ex/0407019; 
and Slivova J 2003 PhD thesis Charles University, Prague
\bibitem{wa80}WA80 Collaboration Albrecht R {\it et al.} 1998 \EPJ{\bf 5} 255 
\bibitem{vitev_gyulassy}Vitev I and Gyulassy M 2002 \PRL {\bf 89} 252301; 
and Vitev I 2004 J.\ Phys.\ G {\bf 30} S791 
\bibitem{adrees}Drees A 2004 private communication
\bibitem{phenix_hipt_62}PHENIX Collaboration Busching H, these Proceeds., {\it Preprint} nucl-ex/0410002 
\bibitem{phobos_hipt_62}PHOBOS Collaboration Back B B {\it et al.} 2004, {\it Preprint}  nucl-ex/0405003 
\bibitem{star_hipt_62}STAR Collaboration Xu Z B, {\it Preprint}  nucl-ex/0411001 
\bibitem{busser73}Busser F W {\it et al.} 1973 Phys.\ Lett.\ B {\bf 46} 471 
\bibitem{busser74}Busser F W 1975 {\it et al.}, Phys.\ Lett.\ B {\bf 55} 232 
\bibitem{eggert75}Eggert K {\it et al.} 1975 Nucl.\ Phys.\ B {\bf 98} 49 
\bibitem{busser76}Busser F W {\it et al.} 1976 Nucl.\ Phys.\ B {\bf 106} 1 
\bibitem{clark78}Clark A G {\it et al.} 1978 Phys.\ Lett.\ B {\bf 74} 267 
\bibitem{angelis78}CCOR Collaboration Angelis A L S {\it et al.} 1978 Phys.\ Lett.\ B {\bf 79} 505 
\bibitem{kourkou79}Kourkoumelis C {\it et al.} 1979 Phys.\ Lett.\ B {\bf 83} 257, 
and 1979 Phys.\ Lett.\ B {\bf 84} 271 
\bibitem{kourkou80}C.~Kourkoumelis C {\it et al.} 1980 Z.\ Phys.\ C {\bf 5} 95 
\bibitem{angelis89}CMOR Collaboration Angelis A L S {\it et al.} 1989 Nucl.\ Phys.\ B {\bf 327} 541 
\bibitem{akesson89}AFS Collaboration T.~Akesson {\it et al.} 1990 Sov.\ J.\ Nucl.\ Phys.\  {\bf 51} 836  
[Yad.\ Fiz.\ {\bf 51} 1314] 
\bibitem{alper75}BSC Collaboration Alper B {\it et al.} 1975 Nucl.\ Phys.\ B {\bf 100} 237 
\bibitem{banner77}Banner M {\it et al.} 1977 Nucl.\ Phys.\ B {\bf 126} 61 
\bibitem{drijard82}CDHW Collaboration Drijard D {\it et al.} 1982 Nucl.\ Phys.\ B {\bf 208} 1 
\bibitem{akesson82}AFS Collaboration Akesson T {\it et al.} 1982 Nucl.\ Phys.\ B {\bf 209} 309 
\bibitem{breakstone95}CDHW Collaboration Breakstone A {\it et al.} 1995 Z.\ Phys.\ C {\bf 69} 55 
\bibitem{axel}Drees Axel 2004, PHENIX Analysis Note (unpublished)
\bibitem{vogelsang_pi0}Aversa F {\it et al.} 1989 Nucl.\ Phys.\ B {\bf 327} 105, 
Jager B {\it et al.} 2003 Phys.\ Rev.\ D {\bf 67} 054005, 
and Vogelsang W (private communication).
\bibitem{diakonou79}Diakonou M {\it et al.} 1979 Phys.\ Lett.\ B {\bf 87} 292 
\bibitem{diakonou80}Diakonou M {\it et al.} 1980 Phys.\ Lett.\ B {\bf 91} 296 
\bibitem{vogelsang_gamma}Gordon L E and Vogelsang W 1993 Phys.\ Rev.\ D {\bf 48} 3136, 
1994 Phys.\ Rev.\ D {\bf 50} 1901, 
and Vogelsang W (private communication).
\bibitem{albajar90}UA1 Collaboration Albajar C {\it et al.} 1990 Nucl.\ Phys.\ B {\bf 335} 261 
\bibitem{kkp}Kniehl B A, Kramer G and Potter B. 2001 Nucl. Phys. {\bf B 597}, 337 
\bibitem{kretzer}Kretzer S 2000 \PRD{\bf 62}, 054001 
\bibitem{bourre}Bourrely C and Soffer J, 2004 Eur.\ Phys.\ J.\ C {\bf 36} 371, {\it Preprint} hep-ph/0311110
\bibitem{resumm}Laenen E, Sterman G and Vogelsang W 2001 {\it Phys. Rev.} D {\bf 63} 114018 
\bibitem{e706_kt}E706 Collaboration Apanasevich L {\it et al.} 1998 {\it Phys. Rev. Lett.} {\bf 81} 2642 
\bibitem{Eggert75}Eggert K {\it et al.} 1975, Nucl.\ Phys.\ B {\bf 98} 93 
\bibitem{Amaldi76}Amaldi U {\it et al.} 1977, Phys.\ Lett.\ B {\bf 66} 390 
\bibitem{Amaldi78}Amaldi U {\it et al.} 1978, Nucl.\ Phys.\ B {\bf 145} 367 
\bibitem{Baksay78}Baksay L {\it et al.} 1978, Nucl.\ Phys.\ B {\bf 141}  1 [Erratum-ibid. 1979 \ B {\bf 148} 538]
\bibitem{Ambrosio82}CERN-Naples-Pisa-Stony Brook Collaboration Ambrosio M {\it et al.} 1982 Phys.\ Lett.\ B {\bf 113} 347
\bibitem{Carboni84}Carboni G {\it et al.} 1985 Nucl.\ Phys.\ B {\bf 254} 697 
\bibitem{Amos85}Amos N {\it et al.} 1985 Nucl.\ Phys.\ B {\bf 262} 689 
\bibitem{vitev04}Vitev I 2004 {\it Preprint} nucl-th/0404052 
\bibitem{adil_gyulassy04}Adil A and Gyulassy M 2004 Phys.\ Lett.\ B {\bf 602} 52 
\bibitem{wang04}Wang X N 2004 Phys.\ Rev.\ C {\bf 70} 031901 
\bibitem{dainese04}Dainese A, Loizides C and Paic G 2004 {\it Preprint} hep-ph/0406201 
\bibitem{phot_reps}See e.g. Peitzmann T and Thoma M H 2002 Phys.\ Rept.\  {\bf 364} 175; 
and Gale C and Haglin K L 2003 in ``Quark Gluon Plasma Vol 3'', World Scientific, Singapore, 
{\it Preprint} hep-ph/0306098 
\bibitem{rasanen_sps_rhic}Rasanen S S 2003 Nucl.\ Phys.\ A {\bf 715} 717 
\bibitem{turbide}Turbide S, Rapp R and Gale C 2004 Phys.\ Rev.\ C {\bf 69} 014903 
\bibitem{dente_peressou}d'Enterria D and Peressounko D, in preparation.
\bibitem{arnold}Arnold P, Moore G D and Yaffe L G 2001 JHEP {\bf 0112} 009 
\bibitem{dumitru}Dumitru A {\it et al.} 2001 Phys.\ Rev.\ C {\bf 64} 054909 
\bibitem{eks98}Eskola K J, Kolhinen V J and Salgado C A 1999 Eur.\ Phys.\ J.\ C {\bf 9} 61 
\bibitem{jeon02}Jeon J, Jalilian-Marian J and Sarcevic S 2003 Nucl.\ Phys.\ A {\bf 715} 795 
\bibitem{arleo04} Arleo F 2004 hep-ph/0406291 
\bibitem{okada04} PHENIX Collaboration Okada K 2004 {\it Proceeds. SPIN'04}

\end{thebibliography}
\end{document}